\documentclass[apj]{emulateapj}

\usepackage{amsmath}
\usepackage{color}
\usepackage{needspace}

\newcommand{\stdev}{{\sigma_{\rm st\, dev}}}
\newcommand{\angstrom}{\textup{\AA}}
\newcommand{\ionizationrecombination}{ionization/recombination\hskip 2pt}
\newcommand{\grav}{{g_\odot^{}}}
\newcommand{\aL}{{a_L^{}}}
\newcommand{\aLz}{{a_{L_{z}}^{}}}
\newcommand{\apres}{{a_p^{}}}
\newcommand{\apresz}{{a_{p_{z}}^{}}}
\newcommand{\LP}{{L_P}}
\newcommand{\LB}{{L_B}}
\newcommand{\popstandard}{A}
\newcommand{\popcold}{C}
\newcommand{\popcorona}{D}
\newcommand{\popdometohot}{B}
\newcommand{\Halpha}{{H$\alpha$}}
\newcommand{\CaII}{{\ion{Ca}{2}}}
\newcommand{\Heteneightthirty}{{\ion{He}{2}  $10830$\,$\angstrom$}}
\newcommand{\Hethreeofour}{{\ion{He}{2}  $304$\,$\angstrom$}}
\newcommand{\Tg}{{T}}

\shorttitle{Cool surge following flux emergence}
\shortauthors{N\'obrega-Siverio, D. et al.}

\begin{document}

%
\title{The cool surge following flux emergence in a radiation-MHD experiment}

%
\author{D. N\'obrega-Siverio \altaffilmark{1,2}, F. Moreno-Insertis\altaffilmark{1,2}, and J. Mart\'inez-Sykora\altaffilmark{3,4}}
\affil{$^1$ Instituto de Astrofisica de Canarias, Via Lactea, s/n, E-38205 La Laguna (Tenerife), Spain}
\affil{$^2$ Department of Astrophysics, Universidad de La Laguna, E-38200 La Laguna (Tenerife), Spain}    
\affil{$^3$ Lockheed Martin Solar and Astrophysics Laboratory, Palo Alto, CA 94304, USA}
\affil{$^4$ Bay Area Environmental Research Institute, Petaluma, CA, USA}

\email{dnobrega@iac.es, fmi@iac.es, juanms@lmsal.com}

%

\begin{abstract}
  Cool and dense ejections, typically \Halpha\ surges,
  often appear alongside EUV or X-Ray coronal jets as a result of the
  emergence of magnetized plasma from the solar interior. Idealized numerical
  experiments explain those ejections as being indirectly associated with the
  magnetic reconnection taking place between the emerging and preexisting
  systems. However, those experiments miss basic elements that can
  importantly affect the surge phenomenon. In this paper we study the cool
  surges using a realistic treatment of the radiation transfer and material
  plasma properties. To that end, the Bifrost code is used, which has
  advanced modules for the equation of state of the plasma, photospheric and
  chromospheric radiation transfer, heat conduction and optically thin
  radiative cooling. We carry out a 2.5D experiment of the emergence of
  magnetized plasma through (meso)granular convection cells and the low
  atmosphere to the corona. Through detailed Lagrange tracing, we study the
  formation and evolution of the cool ejection and, in particular, the role
  of the entropy sources: this allows us to discern families of evolutionary
  patterns for the plasma elements. In the launch phase many elements suffer
  accelerations well in excess of gravity; when nearing the apex of their
  individual trajectories, instead, the plasma elements follow
  quasi-parabolic trajectories with acceleration close to $\grav$. We show
  how the formation of the cool ejection is mediated by a wedge-like structure
  composed of two shocks, one of which leads to the
  detachment of the surge from the original emerged plasma dome.
\end{abstract}


%

\Needspace{5\baselineskip}
\section{Introduction}\label{sec:introduction}
Cool, chromospheric-temperature ejections are key dynamical elements of the
solar atmosphere. Surges, in particular, usually appear in connection with
magnetic flux emergence episodes, in which they are often associated with
hot, high-speed EUV or X-ray jets. Even though observationally known for
several decades now, understanding of the surges has progressed slowly and
there are still many unsolved questions. First detections of chromospheric
surges date back to the 1940s, when they were described as
\Halpha\ absorption markings related with bright eruptions (flares)
corresponding to outward velocities followed by inward motion
\citep{Newton1942,Ellison1942}. Further observational properties were
obtained in the 1970s and 1980s (\citealp{Kirshner1971}, \citealp{roy1973},
\citealp{Cao1980}, \citealp{Schmieder1984}, among others): the surges were
seen as blue and red shifted absorptions in \Halpha\ that have a length of,
typically, $10-50$ Mm, and line-of-sight velocities of a few to several tens
of km s$^{-1}$, reaching, in extreme cases, $100-200$ Mm and $200$ km
s$^{-1}$ respectively. The surges were also observed in
\CaII\ \citep{Rust1976}; a close relationship between \Halpha\ surges and EUV
ejections was found as well \citep{Schmahl1981}.  Later, different
observations focused on the role of the magnetic field, suggesting that the
\Halpha\ surges could be an indirect result of flux emergence processes and
the interaction (possibly reconnection) of the upcoming magnetized plasma
with the ambient coronal field (\citealp{Kurokawa1993},
\citealp{Schmieder1995}, \citealp{canfield1996}, \citealp{Chae1999}). Those
suggestions were based mainly on the detection of the cool ejections next to
emerging bipolar regions and quasi-simultaneously with hot coronal plasma
jets (observed in the EUV or in X-rays).  The high resolution observations of
the past decade \citep[e.g.,][]{Yoshimura2003,Jibben2004, Brooks2007,
  Jiang2007, Uddin2012, Vargas2014} have provided further evidence for the
frequent relation between magnetic flux emergence, chromospheric ejections
and hot jets.  Other chromospheric-temperature ejections such as
macrospicules show some analogies with the surges: they are multithermal
structures observed mainly in \Hethreeofour\ and \Halpha, with a cool core
surrounded by a thin sheath of $1-2\times10^5$~K \citep[e.g.,][]{Bohlin1975,
  Habbal1991, Pike1997, Madjarska2006, Bennett2015}.

Concerning the theoretical effort, the seminal paper by
\citet{Heyvaerts1977} (see also \citealt{forbes1984})
discussed how the emergence of magnetized plasma from the
solar interior could lead to a conflict of magnetic
orientation with the preexisting coronal field and hence to
reconnection and the ejection of hot plasma. Using this flux
emergence paradigm, \citet{Shibata1992a} and
\citet{Yokoyama:1995uq,Yokoyama:1996kx} then showed, through
a 2.5D numerical model with initial uniform coronal field,
that cool plasma could be ejected next to a hot jet as a
consequence of the emergence of magnetic flux from the
interior: the authors tentatively identified those cool
ejections with \Halpha\ surges and described them as resulting from {\it
the sling-shot effect due to reconnection, which produces
a \it whip-like motion} \citep{Yokoyama:1996kx}. 
Their cool surge had density around $10^{-11}$ g cm$^{-3}$,
speeds in the range $\approx 50 - 90$ km
s$^{-1}$, and maximum vertical size of several Mm. 
Nonetheless, due to the computational limitations
of the time, the corona used in the experiment had
unrealistic values of density and temperature. Using the
same sort of setup but with more realistic coronal
parameters, \cite{Nishizuka:2008zl}, through morphological
image comparisons, suggested that the cool ejections
associated with flux emergence could be the cause for
jet-like features seen in \CaII\ H+K
observations. The more recent flux-emergence experiment of
\cite{jiang2012} had a canopy-type configuration of the
ambient coronal magnetic field, and also led to the ejection
of cool and hot plasma. A study in three dimensions of the cool
ejection following magnetic flux emergence has been
published only recently \citep{Moreno-Insertis:2013aa}. This
experiment yielded a cool (from $10^4$~K to a few times
$10^5$~K) and dense (between $10^{-12}$ and $10^{-13}$ g
cm$^{-3}$) wall-like plasma domain surrounding the emerged
flux region. Through Lagrange tracing, the
authors explained the formation of the wall through plasma
which was being transferred from the emerged region attached
to field lines that change connectivity in the main
reconnection site. The cool
ejecta had speeds of typically less than $50$ km s$^{-1}$
and were not collimated.

All those theoretical models, whether 2D or 3D, have been
helpful in providing basic indications for the mechanisms
that may lead to the simultaneous ejection of cold and hot
plasma; nevertheless, they lack essential physical processes
relevant in the photosphere, chromosphere and corona, and
can therefore only be taken as first steps when trying to
understand the physics of the surges. The aim of the
current paper is to provide a new perspective of the cool
ejections introducing some of those physical processes, like
thermal conduction, photospheric and chromospheric radiative
transfer, optically thin radiative cooling and a realistic
equation of state (EOS). To that end we use as computational
tool the Bifrost code \citep{Gudiksen:2011qy}. For a first
approach, in this paper we are using a 2.5D setup. The initial phase of
the flux emergence process takes place through solar-like
granular convection, which influences the sizes of the
resulting structures in the low atmosphere. We can study the
subsequent phenomena of reconnection and plasma ejection in
the atmosphere with high temporal cadence and spatial
resolution, focusing on the formation, maximum development
and decay phases of the surge. The study includes
detailed Lagrange tracing of the mass elements in the surge,
which allows us to analyze in detail their origin and
thermal evolution, the role of the various entropy sources
and the acceleration mechanisms. We show that the cool and
dense ejection is a complex and fascinating phenomenon in
which the entropy sources play an important role.

\begin{figure*}
\epsscale{1.18} \plotone{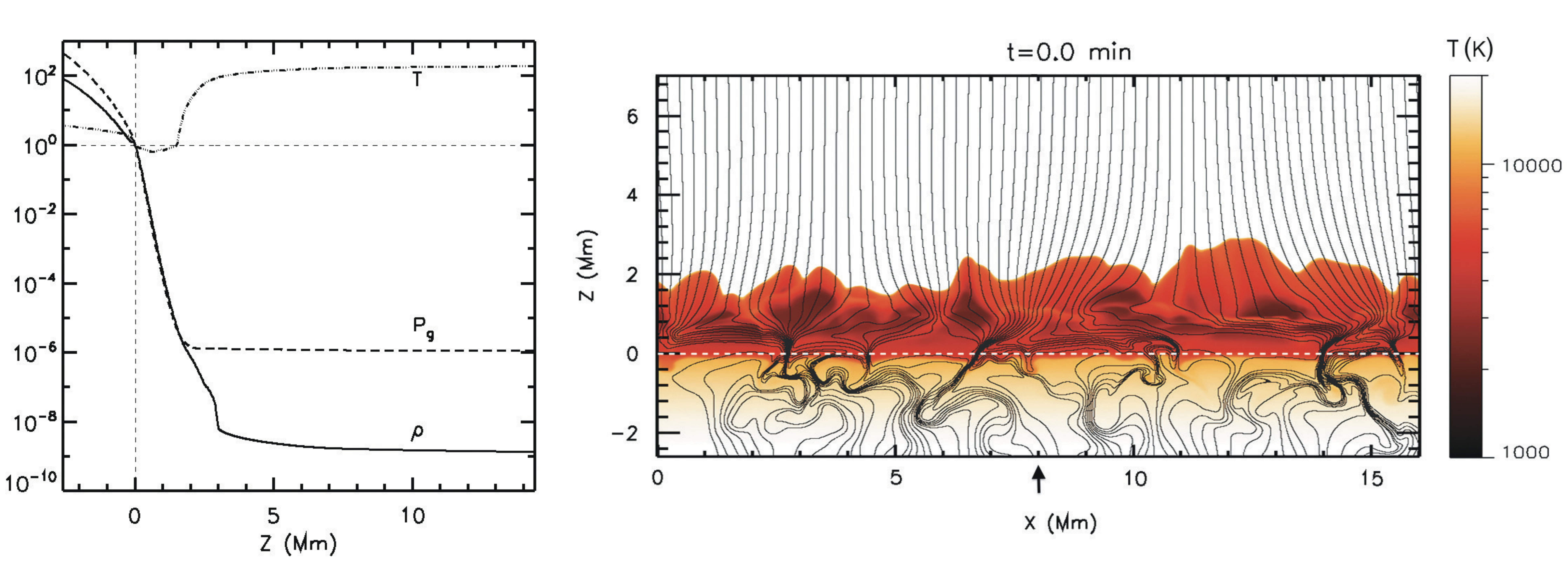}
\caption{Left: Horizontal averages for density $\rho$, gas
  pressure $P_{g}$, and temperature $\Tg$ for the initial
  stratification. The values are normalized to their
  photospheric values at $z = 0$~Mm, namely, $\rho_{ph} =
  3.09 \times 10^{-7}$~g cm$^{-3}$, $P_{g_{ph}} = 1.11
  \times 10^{5}$~erg cm$^{-3}$, and ${\Tg}_{ph}=
  5617$~K. The dotted horizontal line delineates the unity
  and the vertical one, the solar surface. Right:
  Temperature map for the initial background stratification
  for heights between $z=-2.3$~Mm and $z=7.0$~Mm. Magnetic
  field lines appear superimposed in black. The inflow
  region where the tube has been injected is at $x=8$~Mm
  (black arrow). The solar surface is roughly at $z=0$~Mm
  (dashed horizontal white line). \label{fig1}}
\end{figure*}  

The layout of the present paper is as follows. Section
\ref{sec:2} describes the physical and numerical model. In Section
\ref{sec:3} we show the initial phases of the experiment prior to the
initiation of the cool ejection. Sections \ref{sec:cold_wall} and 
\ref{sec:the_cleft} analyze the surge in detail
through its various phases (ejection, detachment and decay), focusing on the heating
sources, kinematics and dynamics of the plasma elements. 
Finally, Section \ref{sec:5} contains the
discussion and the summary.

%

\Needspace{5\baselineskip}
\section{The physical and numerical model}\label{sec:2}
  
\Needspace{5\baselineskip}
\subsection{The numerical code}\label{sec:bifrost}

The experiment we present in this paper has been run using the
radiation-magnetohydrodynamics (RMHD) Bifrost code
\citep{Gudiksen:2011qy}. This code includes thermal conduction along the
magnetic field lines and radiation transfer adequate to the photosphere,
chromosphere and corona; it takes into account entropy sources such as
Spitzer thermal conductivity, optically thin cooling, and radiative losses by
neutral hydrogen, singly-ionized calcium and magnesium, among others; details
are provided in the papers by \cite{Skartlien2000}, \cite{Hayek:2010ac},
\cite{Gudiksen:2011qy}, \cite{Leenaarts:2011qy}, and
\cite{Carlsson:2012uq}. The code also has an equation of state (EOS) that
includes the \ionizationrecombination of the relevant atomic species. On the
other hand, because of the validity range of the radiation tables in the
code, there is an ad-hoc heating term that forces the plasma to stay above
$\Tg = 1660$~K \citep[a discussion in detail concerning this term can be
  found in the paper by][]{Leenaarts:2011qy}. The advantages of the Bifrost
code probably make the simulation in this paper the most realistic one to
date for the formation and dynamics of surges (but see the discussion
concerning various limitations of the present model in
Section~\ref{sec:disc_limitations}).

The description of the model underlying our experiment, is
divided into two parts: (1) the background
stratification, numerical grid and boundary conditions and (2) the twisted
magnetic tube. 

\Needspace{5\baselineskip}
\subsection{Background stratification, numerical grid and boundary conditions}
Concerning the background stratification, we started from a
preexisting statistically stationary magnetoconvection
configuration that includes in a self-consistent manner the
uppermost layers of the solar interior, the photosphere, the
chromosphere, the transition region, and the corona. The
convection patterns range between granular and mesogranular.
The corona has a temperature of about $1$ MK and a
quasi-uniform vertical magnetic field of $10$ G in order to
mimic a coronal hole medium. The initial magnetic field is
contained in the $x-z$ plane, with $z$ being the vertical
coordinate. The left panel in Figure~\ref{fig1} shows the
horizontal averages for our initial condition for density
$\rho$, gas pressure $P_g$ and temperature $\Tg$, all of
them normalized to their photospheric values at $z=0$~Mm,
namely, $\rho_{ph} = 3.09\times10^{-7}$~g cm$^{-3}$,
$P_{g_{ph}} = 1.11 \times 10^{5}$~erg cm$^{-3}$ and
$\Tg_{ph}= 5.62\times10^3$~ K. In the right panel we
present a 2D temperature map where a number of magnetic
field lines have been superimposed in black. In the image,
the granulation pattern is distinguishable through the
vertical field concentrations in the convective downflows
and through the horizontal field lines in the center of the
granules in the photosphere.
   
The physical domain is $0.0$~Mm $\leq x \leq$ $16.0$~Mm and $-2.6$~Mm $\leq z
\leq$ $14.4$~Mm, with $z=0$~Mm corresponding to the solar surface, or more
precisely, to the horizontal level where $< \tau_{500} > = 1$.  The numerical
box has $512\times512$ points in the ($x,z$) directions respectively. The
grid is uniform in the $x$-direction with $\Delta x=31$~km, and non-uniform
in the vertical direction in order to better resolve the lower
photosphere. The vertical grid spacing varies between $19$ km, reached in the
photosphere and chromosphere, and $90$ km at the top and bottom of the
domain. The boundary conditions are periodic in the horizontal direction. At
the top of the box characteristic boundary conditions \citep[as described
  by][]{Gudiksen:2011qy} have been chosen that suppress incoming waves so as
to eliminate any signal reflexion while the plasma can leave the domain.
Additionally, the corona is expected to have temperatures of order $1$ MK but
the two-dimensional nature of this experiment prevents a self-consistent
magnetic heating resulting from photospheric field line braiding, as in the
3D experiment of \citet{Gudiksen:2005lr}. To alleviate this problem, a {\it
  hot-plate} is implemented at the top boundary, meaning a Newton cooling
term that forces the temperature in the boundary cells to stay fixed at
$10^6$ K. For the bottom boundary the code uses a technique often implemented
in magnetoconvection simulations \citep[e.g.,][]{Stein1998, hansteen2007},
namely, it keeps the bottom boundary open so that plasma can go across it,
and constant entropy is set in the incoming material to keep the convection
going, while the rest of the variables is extrapolated.
      
\Needspace{5\baselineskip}
\subsection{The twisted magnetic tube}  
In order to produce magnetic flux emergence, we inject a
twisted magnetic tube with axis pointing in the $y$-direction,
the ignorable coordinate in this 2.5D experiment. The
injection is done through the lower boundary of the box
following the method described by
\citet{Martinez-Sykora:2008aa}. The longitudinal and
transverse components of the magnetic field in the tube have
the canonical form of a Gaussian profile with $r$-independent
pitch (e.g. \citealp{fan2001}),
 
\begin{eqnarray}
	B_y & = & B_0 \, \exp \left( - \frac{r^2}{R_0^2}
        \right), \label{eq:btube1}\\
\noalign{\vspace{3mm}}
	B_{\theta} & = & q \, r \, B_y
        , \label{eq:btube2}\\ \notag
\end{eqnarray}

\noindent where $r$ and $\theta$ are the radial and azimuthal coordinates
relative to the tube axis, $R_0$ is a measure for the tube radius, $q$ a
constant twist parameter, and $B_0$ the magnetic field in the tube
axis. To favor the emergence, we inject the tube in an inflow region, namely
$x_0=8.0$ Mm (marked with a black arrow in the right panel of
Figure~\ref{fig1}). The rest of the parameters are selected within the ranges
that lead to a coherent emergence pattern at the surface; the chosen values
are presented in Table \ref{table1}. The initial axial magnetic flux is
$\Phi_0 = 1.5 \times 10^{19}$~Mx, which is in the range of an ephemeral
active region \citep{Zwaan:1987yf}. This magnetic field configuration 
has positive helicity. The field lines have pitch $\Delta y_p = 2\pi/q = 2.6$~Mm
 independently of the radius. In other words, all field lines execute two turns
  around the axis along a distance of $5.2$ Mm.

\begin{deluxetable}{ccccc}
\tablecolumns{5} \tablewidth{0pc} \tablecaption{Parameters
  of the initial twisted magnetic tube.
\label{table1}} \tablehead{
  \colhead{$x_0$ (Mm)} & \colhead{$z_0$ (Mm)} & \colhead{
    $R_0$ (Mm)} & \colhead{$q$ (Mm$^{-1}$)} & \colhead{$B_0$
    (kG) } } \startdata 8.0 & -2.9 & 0.16 & 2.4 & 19
\enddata
\end{deluxetable}

%

\Needspace{5\baselineskip}
\section{Initial phases}\label{sec:3}

\begin{figure*}
\epsscale{1.27} \centerline{\plotone{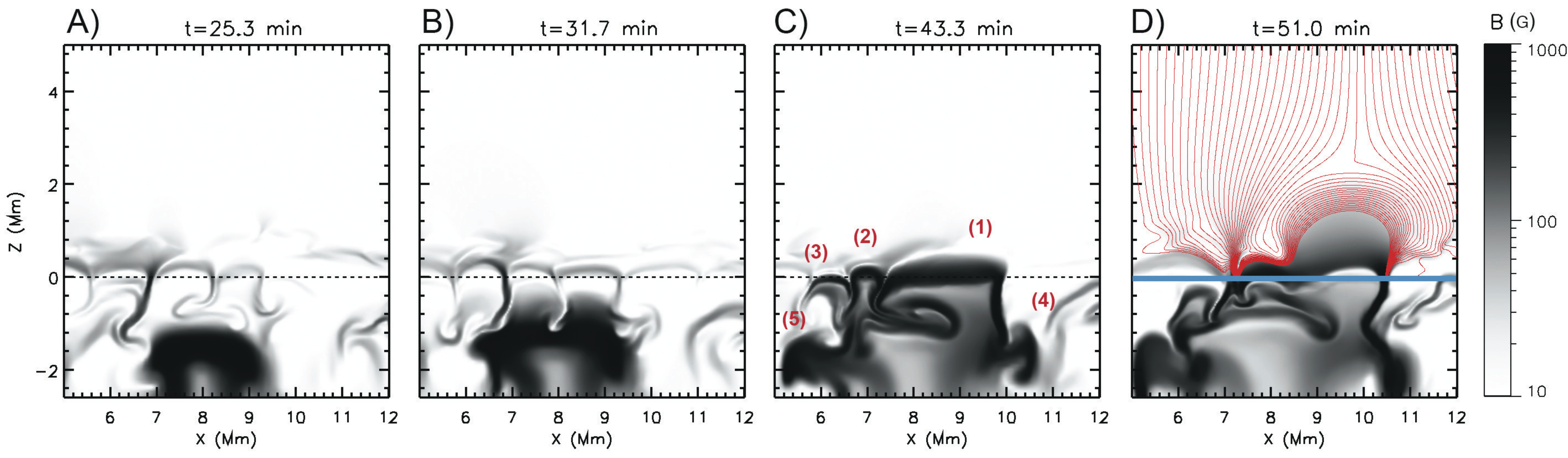}}

\caption{Grey-scale maps of the magnetic field strength for
  the initial phases of the experiment. The horizontal
  dashed line corresponds to the solar surface. In panel C,
  we have indicated with labels 1 through 5 the different
  fragments resulting from the initial tube. Panel D is
  subdivided into two parts, above and below the thick blue
  line: magnetic field lines are superimposed in red, but
  only in the upper subpanel, to avoid blurring the
  structures of the interior.  \label{seccion3}}
\end{figure*}

We call {\it initial phases} the time interval when the
injected magnetic flux is rising through the convection zone
and the photosphere until it reaches the low corona, or more
precisely, until an emerging plasma dome is formed, as
explained in Section~\ref{sec:dome_3}. The initial phases
share similarities with previous RMHD experiments of
magnetic flux emergence (\citealp{Cheung:2007aa},
\citealp{Martinez-Sykora:2008aa}, \citealp{Tortosa2009}) and
also with more idealized MHD experiments
(\citealp{Yokoyama:1996kx}, \citealp{Magara:2001aa},
\citealp{fan2001}, \citealp{archontis2004},
\citealp{Moreno-Insertis:2008ms},
\citealp{Moreno-Insertis:2013aa}).  Figure~\ref{seccion3}
illustrates the module of the magnetic field, $B$, at four
different instants of the initial phases. In the following,
we explain the different panels of the figure.

%

\Needspace{5\baselineskip}
\subsection{Emergence through the convection zone}
The first stage of the rise of the magnetic tube is an expansion away from
the injection point with velocities of order $1$ km s$^{-1}$.  As the tube
rises through the convection zone, it starts to develop a dumbbell shape that
is easily identifiable because of the high field concentration on either side
of the tube axis, panel A in Figure \ref{seccion3}. Afterwards (Panel B), the
action of the convection flows on the rising tube starts to be evident. They
deform and break the twisted magnetic tube into smaller fragments in the
regions of strong shear, typically where the downflows hit the tube.  We can identify
five large fragments during the emergence process. One of the fragments of
the tube, the one tagged ``(1)'' in panel C, reached the surface
approximately $35$~minutes after the initiation of the experiment. At that
instant, this fragment had a horizontal size of 1.3 Mm, i.e., on the order of
a granular size. Two further pieces, ``(2)'' and ``(3)'', get to the surface at
$t \sim$ $38$ and $43$~min respectively, although they are smaller than the
previous one. The fragments labeled ``(4)'' and ``(5)'' were strongly braked
and pushed down by the convection downflows, and they do not reach the
surface. Most of the eruptive phenomena observed in the atmosphere after the
emergence are associated with the first fragment, so we focus attention onto
it in the following.

%

\Needspace{5\baselineskip}
\subsection{Anomalous granulation and buoyancy instability}\label{anomal}
Once in the photosphere, in the transition between super-
and sub-adiabatically stratified regions, the magnetized
plasma starts to pile up, consequently increasing the
magnetic pressure. The enhanced pressure produces a sideways
growth of the fragment leading to an anomalous granule of
about $2.6$ Mm horizontal extent (Figure~\ref{seccion3},
panel C, $x=7.4$ to $10$ Mm), which is twice the size it had
when it reached the surface. Similar anomalous granulation
related with flux emergence was found in the numerical
experiments by \citet{Cheung:2007aa},
\citet{Martinez-Sykora:2008aa}, and \citet{Tortosa2009}, and
in the observations by \cite{orozco2008},
\cite{Guglielmino:2010lr}, and \cite{ortiz2014}, among others.
The later evolution of the anomalous granule occurs in the
frame of the buoyancy instability
\citep{Newcomb:1961aa}, along the general lines described
by \cite{Tortosa2009} and, to some extent, 
also in idealized models without radiation 
\citep[see][]{Magara:2001aa, archontis2004, moreno-insertis_flux_2006, 
murray_etal_2006}. In all those cases the development of the
instability allows the magnetized plasma to rise well above the photospheric 
heights.

%

\Needspace{5\baselineskip}
\subsection{The emerged magnetized dome}\label{sec:dome_3}

Following the buoyancy instability, the plasma belonging to
the anomalous granule suffers a rapid expansion into the
atmosphere with radial velocities of $15$ km s$^{-1}$ at
heights around 1 to 2 Mm. This expansion leads to the
classical dome (or mountain) formation already found in the
past (see, e.g., \citealp{Yokoyama:1996kx},
\citealp{archontis2004}, \citealp{Moreno-Insertis:2013aa}
and references therein). In panel D of Figure~\ref{seccion3}
we show the early stages of the emerged dome, between
$z=0$~Mm and $z=2$~Mm approximately. In the upper subpanel
(above $z=0$, marked by a thick horizontal line in blue),
magnetic field lines are shown superimposed in red: they are
seen to collect into compact field line bunches at the
location of photospheric downflows. One of those bunches is
located at $x \approx 7.2$~Mm; the region between those
lines and the left side of the dome corresponds to a current
sheet that is described in Section~\ref{sec:current_sheet}.

As the dome expands, and as expected for expansion phenomena in the
chromosphere \citep[e.g.][]{Hansteen+DePontieu2006, Martinez-Sykora:2008aa,
  Tortosa2009, Leenaarts:2011qy}, the plasma temperature decreases
significantly, reaching the lower limit allowed in our simulation explained
in Section \ref{sec:2}.  Simultaneously, the dome interior suffers a draining
process owing to the gravitational flows that take place along the loop-like
magnetic field lines, as described by \citet{Moreno-Insertis:2013aa}. The
combination of the expansion and the draining produces a density change from
the values during the first stages of the dome evolution ($10^{-11}$ to
$10^{-12}$ g cm$^{-3}$) to values on the order of $10^{-14}$ g cm$^{-3}$ in
later phases.

%

\Needspace{5\baselineskip}
\subsection{The current sheet: unsteady reconnection}\label{sec:current_sheet}

\begin{figure}
\epsscale{1.21} \plotone{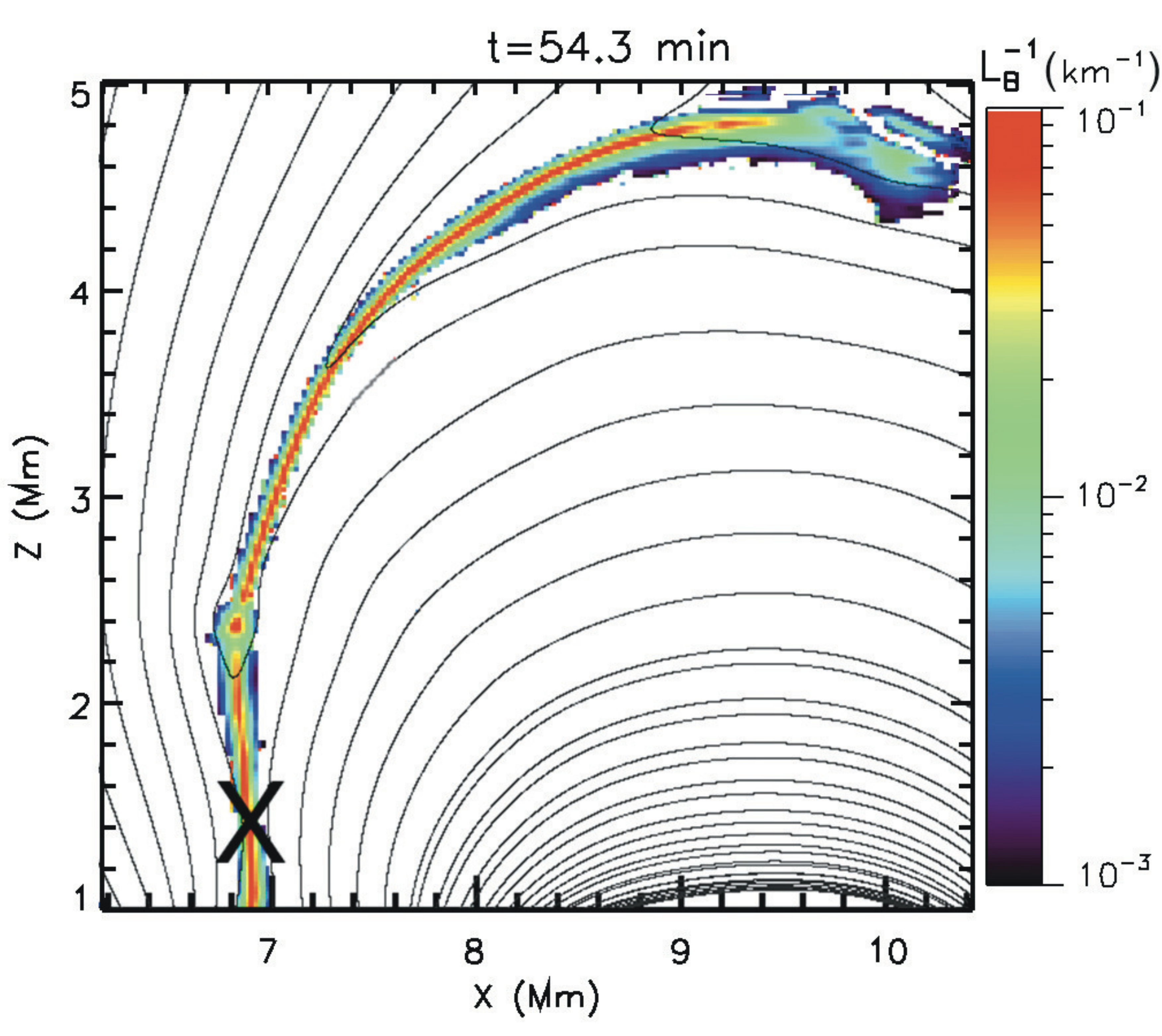}
\caption{Map of the inverse characteristic length of the
  magnetic field $\LB$, see Equation (\ref{eq:lb}),
  illustrating the thin current sheet at the boundary
  between emerged plasma and the corona. Only the pixels
  where $\LB < 1000\ \hbox{km}$ are shown in color. Magnetic
  field lines appear superimposed as solid lines. The black
  cross is the central part of the reconnection
  site. \label{lb_current}}
\end{figure}

The expansion of the dome pushes its magnetic field against
the preexisting vertical coronal magnetic field. This
generates an orientation conflict on the left-hand side of
the dome, giving rise to a thin concentrated current sheet.
In Figure~\ref{lb_current}, we illustrate the latter using
as inverse characteristic length of the magnetic field
variation the quantity
\begin{equation}
	\LB^{-1} = \frac{ \left| \nabla \times
          \hbox{\textbf{\textit{B}}} \right| }{ \left|
          \hbox{\textbf{\textit{B}}} \right|}.
	\label{eq:lb}
\end{equation}
This quantity permits good visualization of any abrupt change of $B$. For
comparison, in a pure rotational discontinuity of the field
$\LB^{-1}$ is $\pi$ times the inverse width of the sheet while in 
a Harris sheet it goes through infinity at the center
of the sheet. In the figure, the pixels where $\LB <
1000\ \hbox{km}$ are shown in color, with magnetic field lines superimposed
as solid lines. There is a reconnection site located at $x\sim6.8$~Mm and
$z\sim1.8$~Mm (black cross). Along its lifetime, the current sheet repeatedly
experiences the formation of plasmoids (like the one at $z\sim2.4$~Mm in
Figure~\ref{lb_current}) through the development of the
tearing-mode-instability \citep{Furth:1963aa}. This behavior has been
detected in previous flux emergence experiments, e.g., in 2D by
\citet{Yokoyama:1996kx}, and in 3D by \citet{archontis2006},
\citet{Moreno-Insertis:2013aa} and
\citet{archontis2014}. In our case, the timescale of
  plasmoid formation is between several tens of seconds and a
  few minutes. This range is compatible with the theoretical value for
  the growth time of the tearing mode \citep[see][]{Goldston1995}: the latter 
  is close to the geometric mean between $\tau_a = L_B/v_a$ and $\tau_d
  = L_B^2/\eta$, where $v_a$ is the Alfven velocity and $\eta$ is the
  diffusivity. In our current sheet, $\tau_a = 10^{-2} - 10^{-3}$ s, and
  $\tau_d = 10^{7} - 10^{8}$ s, and the geometric mean of those quantities is near the
  measured values in the experiment.

As time goes on, the area where the orientation conflict is
located grows in length because of the dome
expansion. During this phase, the plasmoids are ejected as part of the reconnection process,  
probably through the melon seed ejection mechanism \citep{schluter1957}, 
which can be launched when there is an imbalance in the Lorentz force holding 
the plasmoid on its sides along the current sheet. Some of the plasmoids also
merge forming bigger ones as a result of the coalescence
instability \citep{Finn:1977aa}.

%

\begin{figure*}
\epsscale{1.2} \centerline{\plotone{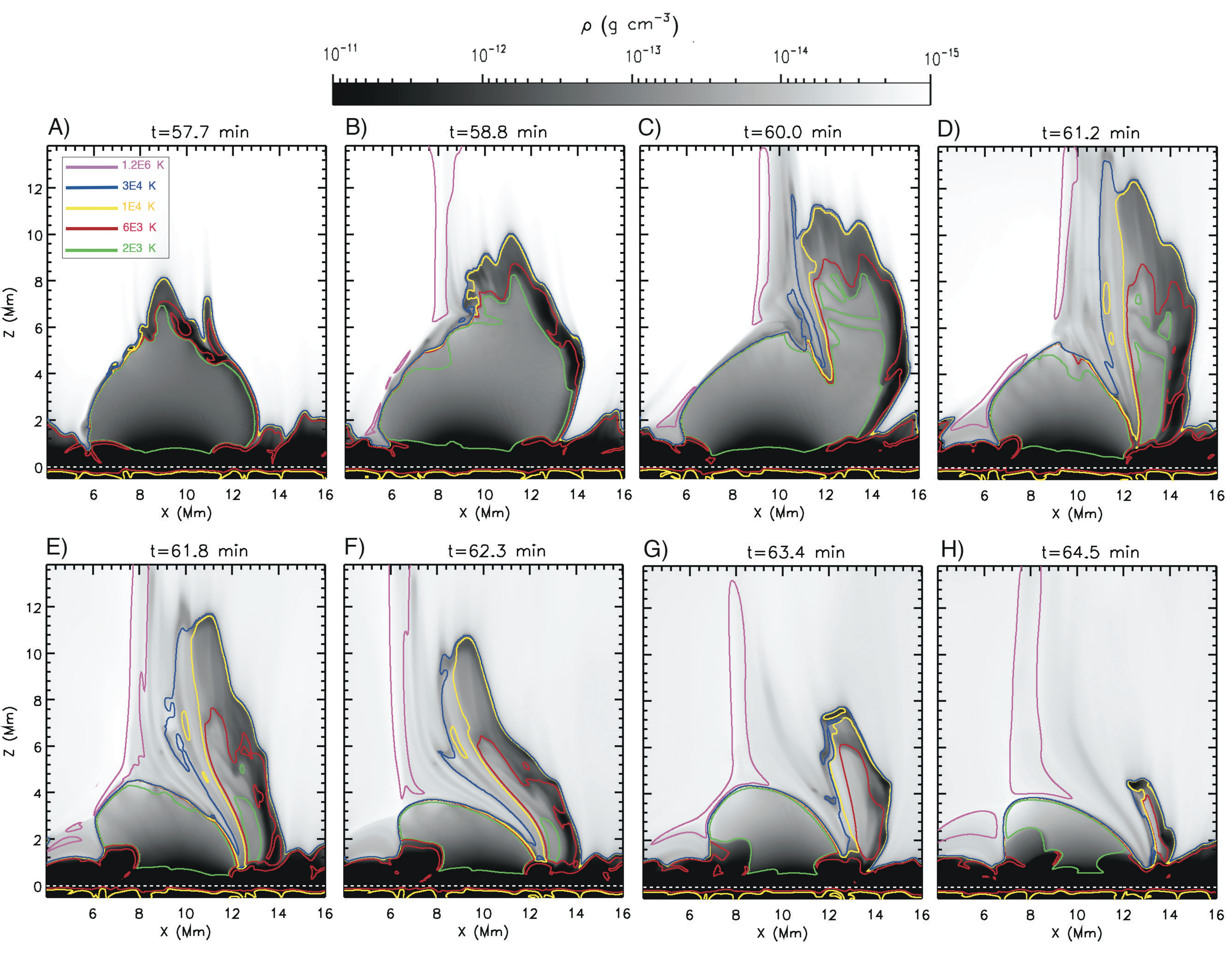}}
\caption{Density maps showing the formation and descending
  phases of the cool and dense ejection. Temperature
  contours have been superimposed to the maps to complete
  the image: the association of colors to temperatures is
  given in panel A. The hot coronal jet is also visible on
  the left-hand side of the dome delineated by the pink
  contours.  See also the accompanying Movie 1.\label{wall}}  
\end{figure*}

\Needspace{5\baselineskip}
\section{The ejection of cool, high-density plasma}
\label{sec:cold_wall}

As a result of the reconnection process taking place at the
boundary between emerged dome and coronal material, a
substantial amount of plasma with chromospheric temperatures
and densities is ejected to coronal heights. We refer to
this phenomenon as the {\it cool and dense ejection} or {\it
  surge} and avoid the word {\it jet}, since the ejecta are
not collimated and do not have large speeds, as shown in
this section. In Figure~\ref{wall}, the overall evolution of
the surge from the initial stages to the decay phase is
illustrated using grayscale maps for the density and with a
number of temperature contours with values indicated in
panel A. In panels A and B we can see an apparent peeling
process that is carrying dense and cool plasma to greater
heights toward the right of the dome. At the same time, a
hot coronal jet is forming on the left side as can be
identified through the pink contours in Panel B. In panel C,
the dome seems to be splitting into two parts at $x \approx
12$~Mm. In panel D, we can distinguish the cool and dense
plasma ejection as the elongated structure located to the
right of the emerged dome with temperatures below $3 \times
10^4$~K, i.e., chromospheric temperatures, including a
colder core of lower temperatures down to $2\times 10^3$~K.
Around this instant, the ejecta reach their maximum height,
$z = 13.2$~Mm, and the density range in the surge is between
$10^{-14}$~g cm$^{-3}$ and $10^{-11}$~g cm$^{-3}$. The rest
of the panels (E, F, G and H) show the decay phase. During
the decay, the surge moves first to the left and then to the
right in a swaying motion caused by the Lorentz force
associated with the bending of its magnetic field lines. The cool
surge remains as an easily identifiable feature until $t
\approx 66$~min, so its lifetime can be estimated to be
about $7-8$~minutes. The accompanying Movie 1 shows the time
evolution of the density and temperature of the system. 

\begin{figure*}
\epsscale{1.15} \centerline{\plotone{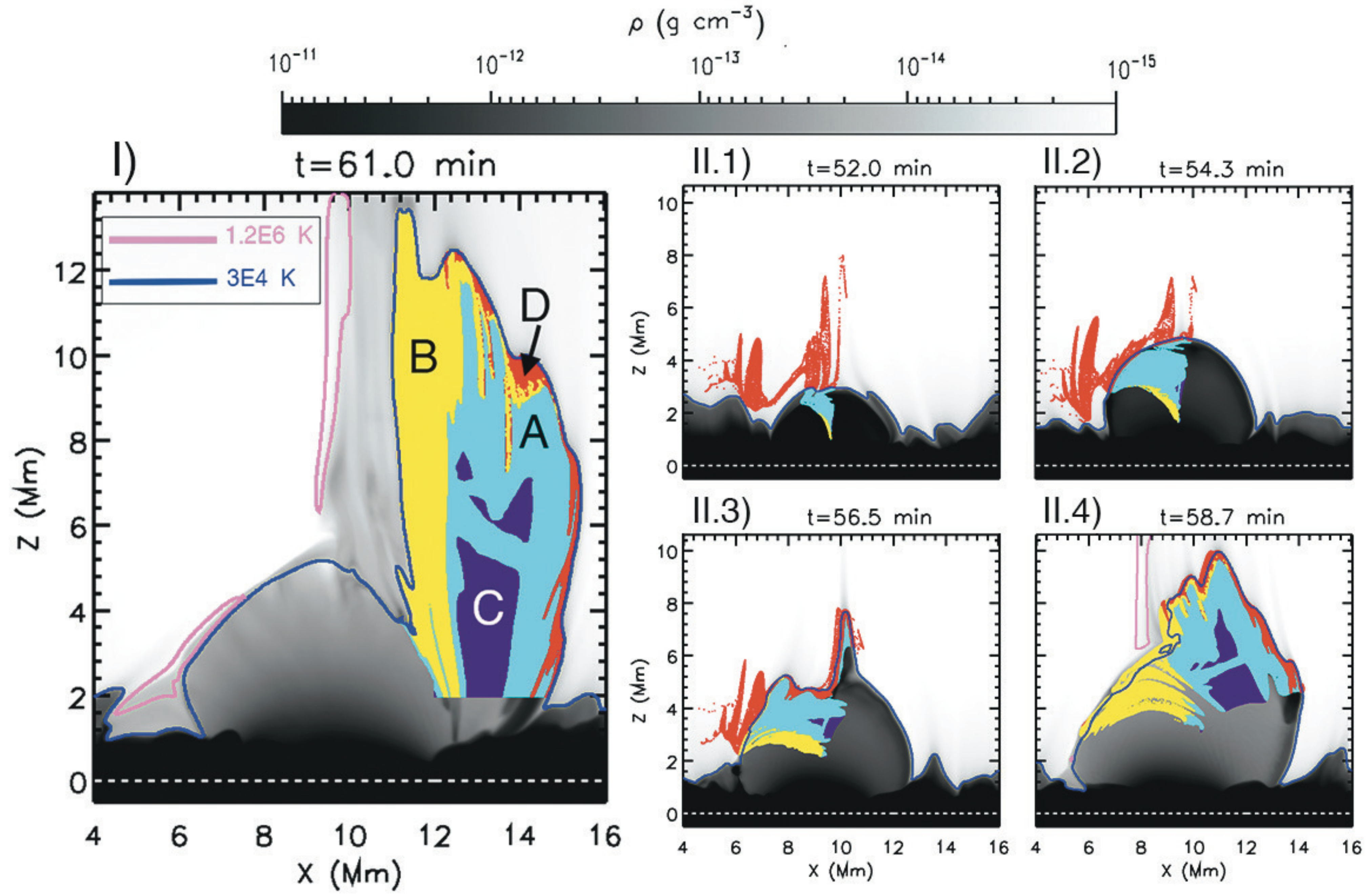}}
\caption{Panel I: Density map showing the basic distribution of the 
more than $3 \times 10^5$ Lagrangian tracers by means of colored domains 
(cyan, yellow, purple and red). The domains correspond to the four 
different populations discussed from Section~\ref{subsec:thermo} onward and are 
tagged with the corresponding capital letters. Panels II.1 through II.4:  Evolution of the ensemble of 
tracers from time $52$ min onward. A more continuous illustration of the time evolution is given in the 
accompanying Movie 2). \label{dome_thermo}}
\end{figure*}

In order to analyze the fundamental aspects of the surge, we
have followed more than $3 \times 10^5$ plasma elements
through Lagrangian tracing. The choice of tracers was
carried out at the time of maximum vertical extent of the
cool ejection, $t= 61$~min, and is shown in panel I of
Figure~\ref{dome_thermo} with dots of different colors
superimposed on the image. We set the side and top boundary
of the surge to coincide with the isocontour $\Tg = 3 \times
10^4$~K (blue curve) and use as lower limit the $z = 2$~Mm
horizontal axis. The tracers are then evenly distributed in that
domain with high-resolution spacing $\delta x = \delta z =
10$~ km. For later reference, we have drawn the
tracers in four different colors (cyan, yellow, purple and
red) according to the four different populations of
plasma elements that are introduced and discussed from
Section~\ref{subsec:thermo} onward. The resolution is high 
enough for the individual tracers to be indistinguishable in the figure: 
the domains looks like a continuous surface. Once the distribution is
established, we follow the tracers backward in time until
$t=51$~min, to study their origin, and also forward, until
$t=65$~min. The tracking has a high temporal cadence of
$0.2$~seconds in order to reach good accuracy even in 
locations with high gradients and phases of fast changes,
like when going through the current sheet.

The rest of this section is divided into three blocks devoted to the heating and cooling sources of the surge 
(Section \ref{subsec:thermo}), the plasma
acceleration (Section \ref{sec:dyn}), and the velocities, densities
and temperatures (Section \ref{sub:velocities}).

%

\Needspace{5\baselineskip}
\subsection{Heating and cooling sources}\label{subsec:thermo}

\begin{figure}
\epsscale{1.20} \centerline{\plotone{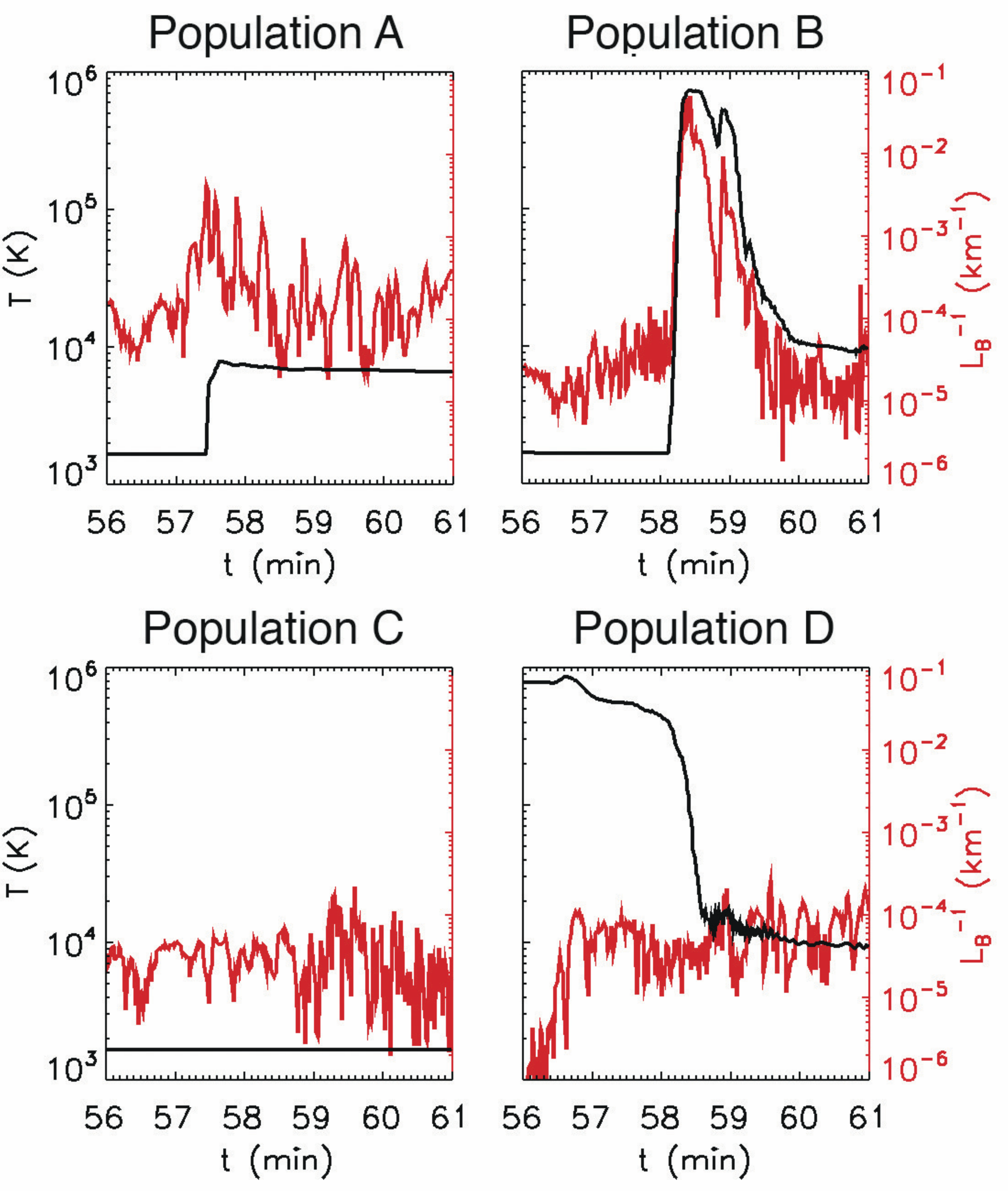}}
\caption{Set of panels showing the time evolution for four
  different plasma elements followed by Lagrangian
  tracing. The chosen elements are representative of the
  four different populations we found concerning the thermal
  properties. In the panels, the temperature $\Tg$ is
  plotted in in black while $\LB^{-1}$ in red. This last
  quantity allows to identify the proximity of the element
  to the current sheet. \label{dynamics_4}}
\end{figure}

Here we analyze the heat sources and sinks in the surge, since they are key
for understanding its structure and evolution. From all the entropy sources
included in the Bifrost code, the relevant ones for the surge are those
resulting from the Spitzer thermal conductivity, the optically thin cooling,
the radiative losses by neutral hydrogen, and the ohmic and viscous
heating. The rest, like those associated with the chromospheric radiative
losses by singly-ionized calcium and magnesium, have much longer
characteristic times and need not be discussed. The following results are
focused on the rising phase of the ejecta until they reach their maximum
height at around $t = 61$~min. Thereafter, the characteristic times of the
heating and cooling processes become much longer than the general
evolutionary timescale of the surge: the temperature changes in the decay
phase are due to adiabatic compression.

The study of the thermal properties of the individual Lagrangian elements
allows one to discern four different plasma populations within the ejecta of
different origin and evolution that we have identified with labels ``\popstandard'',
``\popdometohot'', ``\popcold'' and ``\popcorona'', and drawn in colors cyan,
yellow, purple and red, respectively, in Figure~\ref{dome_thermo}. 
In Panel I we have already introduced the distribution of the Lagrangian 
elements when the cool ejection reaches its maximum height. 
Panels II.1 -- II.4 show the evolution of those elements during previous 
stages of the surge. These panels illustrate the origin of the plasma in the surge 
and how the different populations evolve to give rise to the distribution shown in Panel I. 
A more complete view of the formation of the surge is also provided via the 
accompanying Movie 2. The nature of the 
different populations is analyzed in the following.


\Needspace{5\baselineskip}
\subsubsection{Population \popstandard}\label{sec:popstandard}

Population \popstandard, plotted in cyan in Figure~\ref{dome_thermo},
corresponds to plasma originating in the dome (see panel II.1) and it is heated through 
Joule and viscous dissipation during
the early stages of formation of the surge. Owing to the high density of
this population, those entropy sources are not able to heat the plasma to
values above $3 \times 10^4$~K. At $t=61$~min, this population covers $44\%$
of the cross section of the surge and its total mass per unit length in the
$y$ direction is $10^{-3}$ g Mm$^{-1}$.  The top-left panel in
Figure~\ref{dynamics_4} shows with a black solid line the time evolution of
the temperature of a representative plasma element of this population. The
element jumps from dome temperatures, close to $2\,\times 10^3$~K, to values
around the temperature of the hydrogen \ionizationrecombination, namely
$6\,\times 10^3$~K. In the same panel (red curve), the values of $\LB^{-1}$
at the positions reached by that plasma element indicate that it passes near
the current sheet, but not quite through it: the typical values of $\LB$ in
the current sheet are less than $100$ km (see Figure \ref{lb_current}).


\Needspace{5\baselineskip}
\subsubsection{Population \popdometohot}\label{sec:popdometohot}

\begin{figure}
\epsscale{1.20} \centerline{\plotone{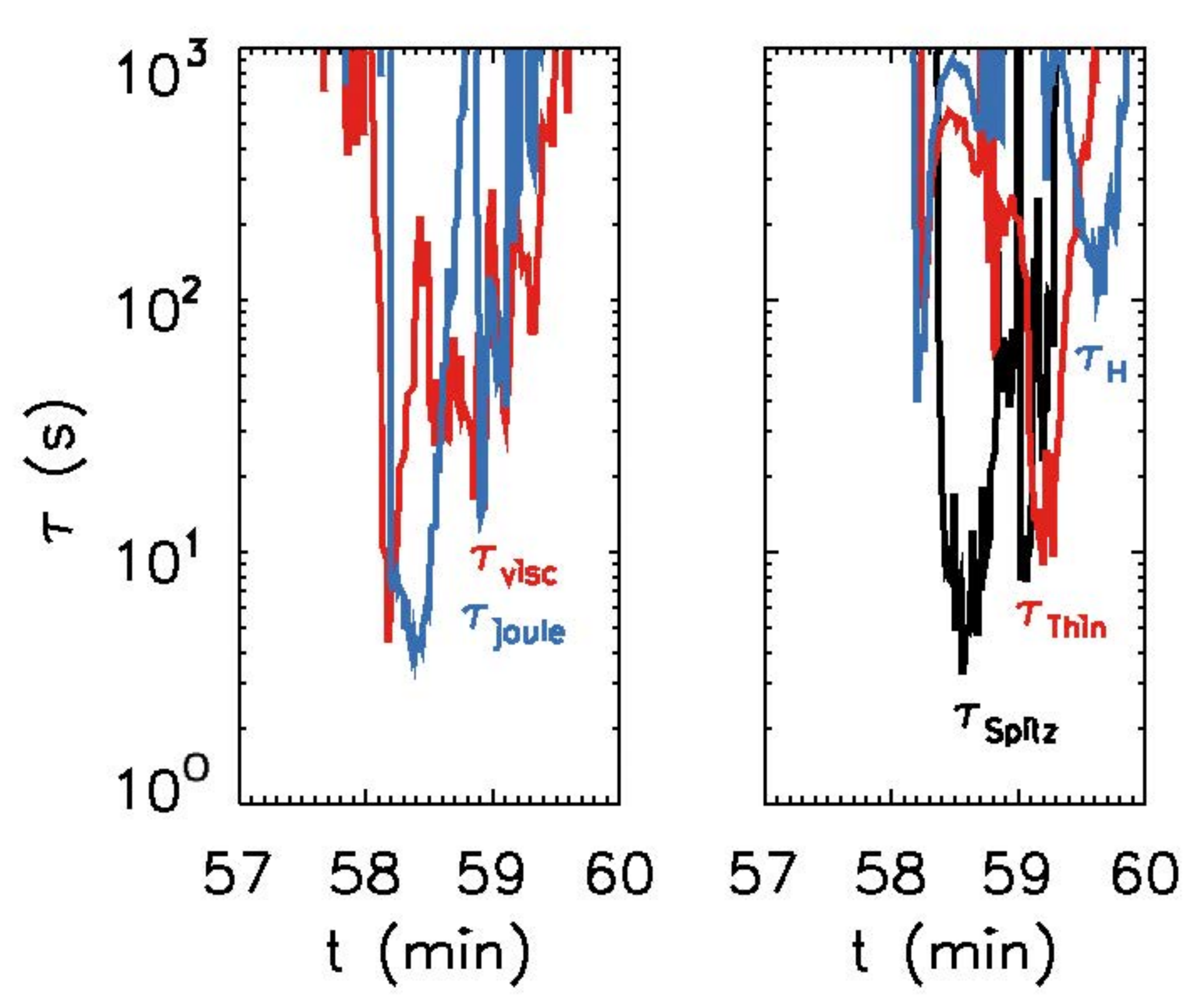}}
\caption{Characteristic times of the entropy sources and
    sinks for the representative plasma element of Population
    \popdometohot~used in Figure~\ref{dynamics_4}. Left panel: Joule 
($\tau_{Joule}^{}$) and viscous ($\tau_{visc}^{}$) heating sources. Right panel: 
    Thermal conduction ($\tau_{Spitz}^{}$), optically thin radiative losses
    ($\tau_{thin}^{}$), and radiative losses by neutral hydrogen
    ($\tau_{H}^{}$). \label{times_popb}}
\end{figure}

The second group of Lagrange tracers is what we call Population
\popdometohot, drawn in yellow in Figure~\ref{dome_thermo}.  Its defining
feature is that the plasma elements in it, in spite of originating in the
dome, reach temperatures between $10^5$ and $10^6$~K during the launch phase,
and then cool down to temperatures below $3 \times 10^4$ K. This family
leaves the dome later than the elements of population \popstandard, as shown
 in the panels II.1 -- II.4 of Figure~\ref{dome_thermo}; it ends
up covering $34\%$ of the surge's cross section at $t=61$~min, but it has
comparatively low densities, so its integrated mass at that time is $4.7\,\times 10^{-5}$
g Mm$^{-1}$ only.  There are two main reasons for the sudden increase in
temperature of the elements of this population, namely:

\begin{enumerate}

\item Unlike for Population \popstandard, some of its plasma elements pass
  through the current sheet, and are strongly heated there (see the yellow tracers
  above the blue temperature contour, $3 \times 10^{4}$ K, in panels II.3 and II.4). The top-right
  panel in Figure~\ref{dynamics_4} depicts an example of this behavior for a
  representative member of this population. The characteristic
  length $\LB$ reaches a small value, around $25$ km, after which it
  decreases, indicating that the plasma element is then leaving the current
  sheet. When in the current sheet, the Joule dissipation and, to a
  lesser extent, the viscous dissipation become highly efficient, with
  short characteristic timescales from several seconds to a few tens of
  seconds, as shown in the left panel of Figure~\ref{times_popb}.   

\item Some plasma elements of this Population, those close to the
  blue contour on the left side of the surge in panel I of Figure~\ref{dome_thermo}, are
  affected by their passage through a strong shock. This shock is a central
  feature of the dynamics of the surge and is described separately
  (Section~\ref{sec:the_cleft}).

\end{enumerate} 

Additionally to the foregoing, the heating processes are particularly
effective at increasing the temperature of the particles of this population
given their comparatively low initial density: their late ejection from the
emerged dome implies that the density of the latter has already been
substantially reduced through the gravitational draining explained in
Section~\ref{sec:dome_3}. The late ejection furthermore explains their
appearance on the left hand side of the surge and with comparatively low
density (see the left side of the surge in panel D of Figure~\ref{wall} in comparison with its right side).

The short duration of the high temperature spurt of these mass elements is
explained by the activation, when the temperature is nearing $10^6$ K, of 
thermal conduction and optically thin radiative losses as effective
entropy sinks. The associated characteristic times ($\tau_{Spitz}^{}$,
$\tau_{thin}^{}$, respectively) for the plasma element studied above can be
seen (Figure~\ref{times_popb}, right panel) to reach low values of several
seconds ($\tau_{Spitz}^{}$) or of a few tens of seconds
  ($\tau_{thin}^{}$). When $T$ decreases to values around $10^4$ K, the
  radiative losses by neutral hydrogen can also be important (see the curve
  labeled $\tau_H^{}$). It is through these cooling processes that the
  elements of this population eventually adopt the cool temperatures of the
  surge.


\Needspace{5\baselineskip}
\subsubsection{Population \popcold}\label{sec:popcold}

Population \popcold, plotted in purple in Figure~\ref{dome_thermo},
is a fraction of the surge coming from the dome that
maintains its initial temperature in the time range
shown. At $t = 61$~min, this population covers $15\%$ of
the cross section of the surge and has a small mass
content, $2.8\,\times 10^{-5}$ g Mm$^{-1}$. 
The Lagrangian tracing shows that the plasma elements were
dragged passively from the dome, following the motion of the
magnetic field lines explained in Section \ref{sec:accel_launch}. 
Along this process, they are never heated by Joule or viscous dissipation. 
The local values of $\Tg$ and $\LB^{-1}$ for a representative plasma element of this population are
given in Figure~\ref{dynamics_4}, bottom-left panel. In fact, those plasma
elements expand along their motion, which explains why this population is barely visible in 
panels II.1 and II.2 of Figure~\ref{dome_thermo} in comparison with panels II.3 and II.4). 
The density of the elements decreases by
approximately one order of magnitude, but their temperature is kept constant
through the ad-hoc heating term mentioned in Sections~\ref{sec:2} and  
\ref{sec:disc_limitations}.


\Needspace{5\baselineskip}
\subsubsection{Population \popcorona}\label{sec:corona}

A small fraction of the Lagrange elements in the surge, plotted in red in Figure
\ref{dome_thermo}, have tracks that start
in coronal heights at $t < 57$ min (panels II.1 -- II.3). The ensemble of such elements is called
Population \popcorona\ in the following. They cover $7\%$ of the cross section of the
surge at $t = 61$ min (panel I). The temperature and $\LB^{-1}$ evolution for a representative plasma
element of this population is shown in the bottom-right panel of
Figure~\ref{dynamics_4}. The tracks start at heights well above the
reconnection site, with standard coronal temperature and density. When
approaching the current sheet, though, these elements go through regions of
large density gradients. The diffusion term included in the mass conservation
equation becomes important, with characteristic timescale less than one
minute, i.e., similar to the evolutionary time of the particles. The
evolution that takes place then is effectively equivalent to a process of
mixing across the density gradient with plasma elements coming from the dome,
after which their behavior is equivalent to that of population
\popstandard. This kind of effective mixing is peculiar of population
\popcorona: the density diffusion term is small for the elements of the other
populations. A proper study of the evolution of this population must
therefore await a numerical experiment with much higher spatial resolution
and correspondingly small numerical diffusion (see also the discussion in
Section~\ref{sec:disc_limitations}).

%

\Needspace{5\baselineskip}
\subsection{The acceleration of the surge}\label{sec:dyn} 

We turn now to the dynamics of the surge and study the acceleration of the
plasma elements first during the launch phase (Section
\ref{sec:accel_launch}) and then when they are near the apex of their
trajectory (Section \ref{sec:accel_apex}).

\Needspace{5\baselineskip}
\subsubsection{Acceleration during the launching phase}\label{sec:accel_launch}

The initial acceleration of the mass elements of the surge takes place when
they are not far from the thin current sheet that covers the top-left region of the dome (Section
\ref{sec:current_sheet}).  In this region, the Lorentz force may reach values
well above gravity because of the high curvature of the magnetic field lines after
reconnection. The gas pressure gradient may also reach large values
because of the low values of the magnetic field at the center of the current
sheet. For an estimate, call $\aL$ and $\apres$ the acceleration associated with those
forces, use $\LB$ as given in Equation~(\ref{eq:lb}) and define $\LP$ as the
corresponding length scale of variation of the gas pressure. One obtains:

\begin{eqnarray}
	\left | \frac{\aL}{\grav} \right |\; &=& \;
        \frac{v_a^2}{2\, L_{B}\, \grav} \;\approx\; 
        18\,\frac{({v_a})_{\rm 100}^2}{(\LB)_{\rm Mm}^{}} ,
	\label{eq:aL}\\
\noalign{\vspace{3mm}}
	\left | \frac{\apres}{\grav} \right | \; &=& \;
           \frac{c_s^2}{\gamma\, \LP\, \grav} \; \approx \; 
        22\,\frac{({c_s})_{\rm 100}^2}{(\LP)_{\rm Mm}^{}} ,
\label{eq:aP}\\ \notag
\end{eqnarray}

\noindent where the subindices ``$100$'' and ``Mm'' indicate
velocities measured in units of $100$ km s$^{-1}$ and lengths measured
in Mm, respectively, and $v_a$, $c_s$, $\grav$ and $\gamma$ have their
customary meaning (Alfven and sound speed, solar gravity, and ratio of
specific heats, respectively).  In the reconnection region, the
characteristic lengths are substantially smaller than $1$ Mm and
either the Alfven velocity or the sound speed (or both) are of order
$100$ km s$^{-1}$. Equations~(\ref{eq:aL}) and (\ref{eq:aP}) tell us,
therefore, that $\aL$ and $\apres$ can easily exceed $\grav\,$; in fact,
in some extreme cases they reach values of a
few times $100\ \grav$ for a short period of time.

\begin{figure}
  \epsscale{1.25} \centerline{\plotone{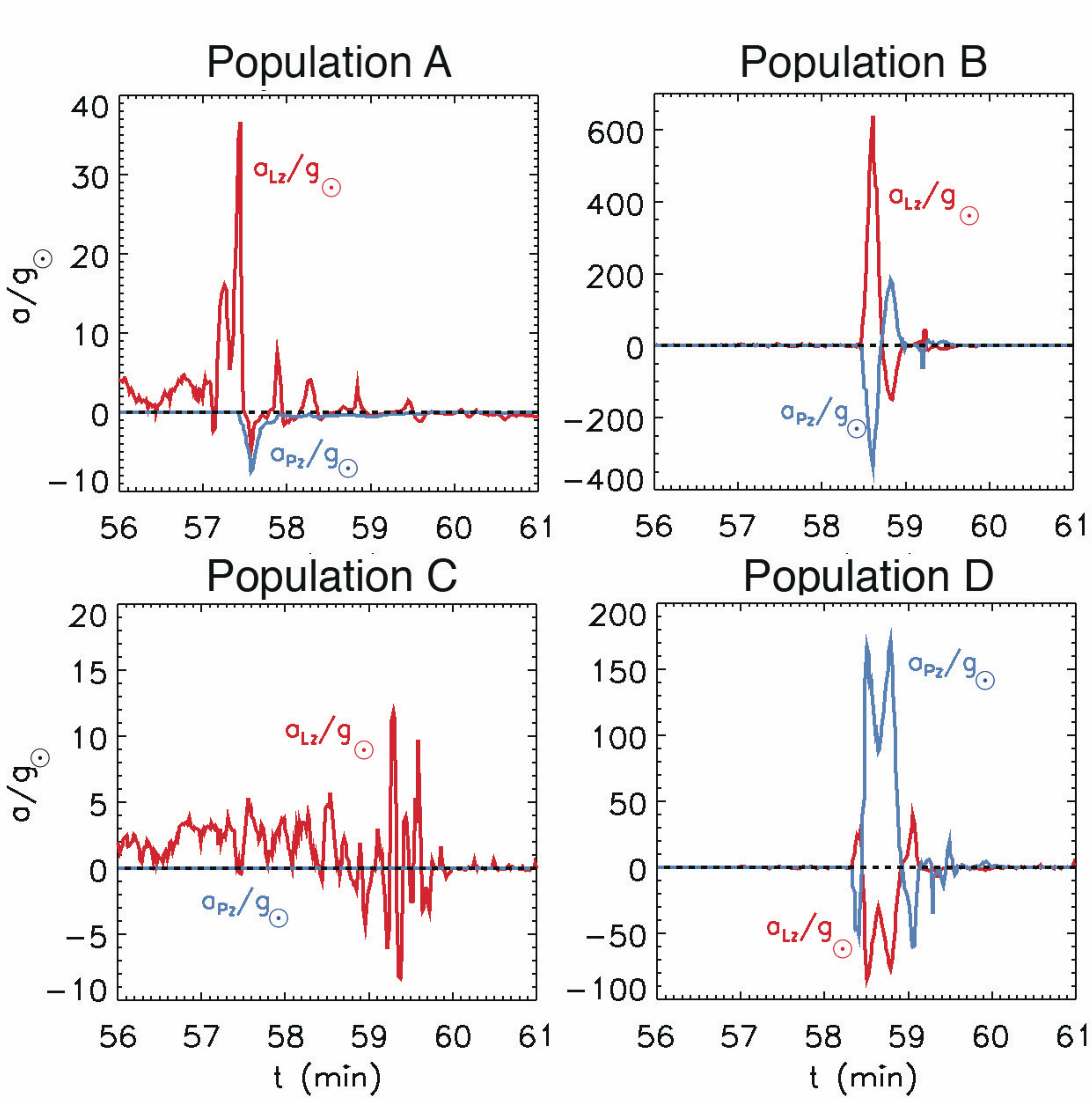}}
\caption{The vertical acceleration components $\aLz$
  and $\apresz$ in units of $\grav$, in red and blue
  respectively, of the same representative plasma elements
  used in Figure~\ref{dynamics_4}. The panels illustrate the
  high acceleration of the plasma during the launch of the
  cool ejection.
\label{new_dynamics}}
\end{figure}

Figure~\ref{new_dynamics} shows the vertical acceleration
components $\aLz$ (red) and $\apresz$ (blue) for the
representative Lagrangian elements used in
Figure~\ref{dynamics_4}. We note the following behavior:

\begin{itemize}

\item In Section~\ref{sec:popstandard} we saw how the elements of population
  \popstandard\ pass near, even though not quite through, the current
  sheet. In the top-left panel of Figure~\ref{dynamics_4} we see how, in that
  phase, they are ejected by the Lorentz force with accelerations of tens of $\grav$. 
  We also note the close relationship between the dynamic and
  thermodynamical changes (compare this panel with the corresponding
  one in Figure~\ref{dynamics_4}).

\item In the case shown in the top-right panel (population
  \popdometohot), the element is ejected upward in a short time
  interval around $t = 58.5$~min. The acceleration values are extreme
  in this case, reaching $\aLz/\grav = 6.4\times 10^2$ and
  $\apresz/\grav = -3.5 \times 10^2$ and last for about $10$~s. Those
  values result from the fact that the element is going through the
  current sheet at that point and the characteristic lengths are
  correspondingly small, $\LB = 40$ km and $\LP = 50$ km (compare
  these lengths with those shown in Figure~\ref{lb_current}).

\item The elements in Population \popcold, like the one shown in
  the left-bottom panel, are the furthest away from the
  current sheet and their characteristic lengths are the
  largest ones. As a consequence, the acceleration values
  are lower than for other populations but are,  anyway, typically a few to
  several times $\grav$. 
  The plasma in this population is dragged by the magnetic field following the highly dynamical motion
initiated in the current sheet and the gas pressure does
not play any important role.

\item The right-bottom panel shows an element from Population
  \popcorona. The large pressure gradients in the boundary between
  the corona and the current sheet lead to a small characteristic
  length $\LP = 100$ km and to the predominance of $\apresz$ compared
  to $\aLz$. The  extreme values of the acceleration in this case are around
  $\aLz/\grav = -0.8\times 10^2$ and $\apresz/\grav = 1.6
  \times 10^2$, and last for about $20$~sec.

\end{itemize}

In the foregoing we have proved that the Lorentz force and gas
pressure gradients in the region at and near the current sheet can
easily cause substantial accelerations of tens to hundreds of $\grav$. This
may seem quite large, but it is naturally associated with 
the fact that the plasma elements must jump by, in some cases, $6$ Mm
in height (from the top of the dome to the top of the resulting surge,
check Figure~\ref{wall}) in a matter of, say, one minute. As an
elementary calculation shows, sustained accelerations of several times
$\grav$ (or impulsive accelerations of from tens to hundreds of
times $\grav$) ought to be expected.

\Needspace{5\baselineskip}
\subsubsection{Acceleration near the apex of the trajectories}\label{sec:accel_apex}

\begin{deluxetable}{ccccc}
\tablecolumns{5}
\tablecaption{Statistical moments of the three
  distributions of
  Figure~\ref{ace_histo}.  \label{table2}} \tablehead{
  \colhead{Curve} & \colhead{$t - t_{apex}$~(min)} &
  \colhead{Mean $/ \grav$} & \colhead{$\stdev / \grav$} 
  &  \colhead{Mode $/ \grav$}
} \startdata
Black & [-1,1] & -0.99 & 6.3 & -1.1 \\

Red & [-2,2] & -0.48 & 6.6 & -1.1 \\

Blue & [-3,3] & \ 0.18 & 11.\hbox to 3mm{\hfill}  & -1.0  \enddata 
\end{deluxetable}

\begin{figure}[h]
\epsscale{1.25} \centerline{\plotone{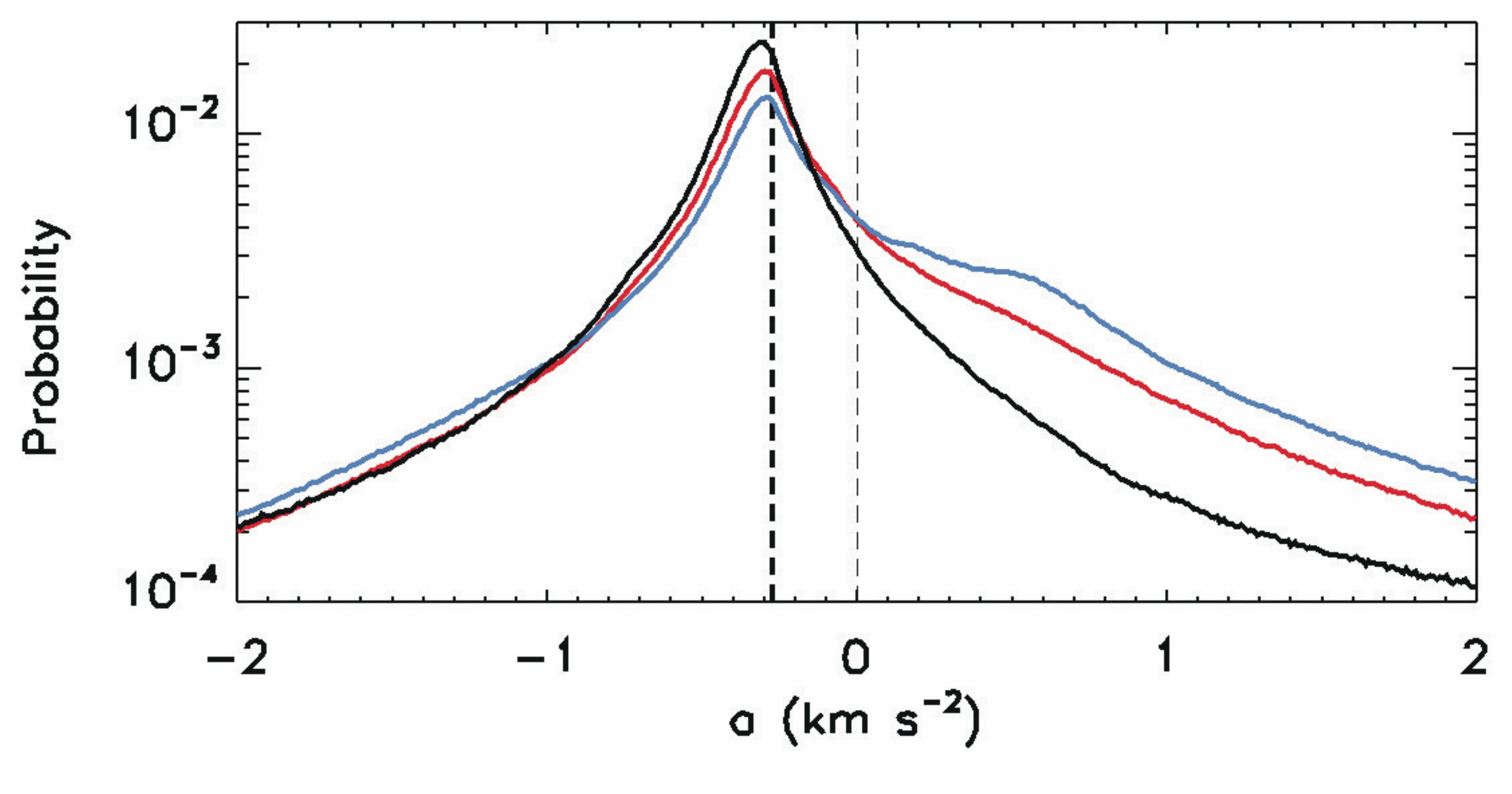}}
\caption{Histograms of the vertical accelerations for a time interval
  around the apex of the trajectories $|\,t - t_{apex}| = 1$~min (black curve), 
  $2$~min (red curve), and $3$~min (blue curve). The vertical
  lines mark $-\grav$ (thick) and the zero acceleration value
  (thin). The statistical properties of these distributions are collected in
  Table \ref{table2}. \label{ace_histo}}
\end{figure}

We examine now the acceleration of the plasma elements
during the central period of development of the surge, namely 
when the Lagrange elements are close to the apex of their
trajectories. To do this, we call 
$t_{apex}$ the time when each individual element reaches its maximum
height and 
use $t - t_{apex}$ as time variable. Figure~\ref{ace_histo} contains
three histograms for the vertical accelerations of the plasma elements
for $|\,t - t_{apex}| = 1$~min (black curve), $2$~min (red curve), and
$3$~min (blue curve). Additionally, we have carried out a statistical
study using the sample of the vertical accelerations of all elements
during the indicated time intervals with a cadence of $0.2$~s.
The basic moments of the statistical distribution and their mode
are given in Table \ref{table2}. 
The three distributions are highly peaked (positive kurtosis) and their most
frequent value (the mode) is very near $-\grav$ in all cases. The mean of the
most representative histogram (black curve) also coincides with
$-\grav$. Yet, the distributions are not narrow, with standard deviations
ranging from $6\,\grav$ to $11\,\grav$. Also: as wider time ranges around $t_{apex}$
are chosen (red and blue curves), upward accelerations linked to the launch
phase are more frequently represented and the mean of the distributions
then shifts toward positive values.

%
\begin{figure}
\epsscale{1.20}
\centerline{\plotone{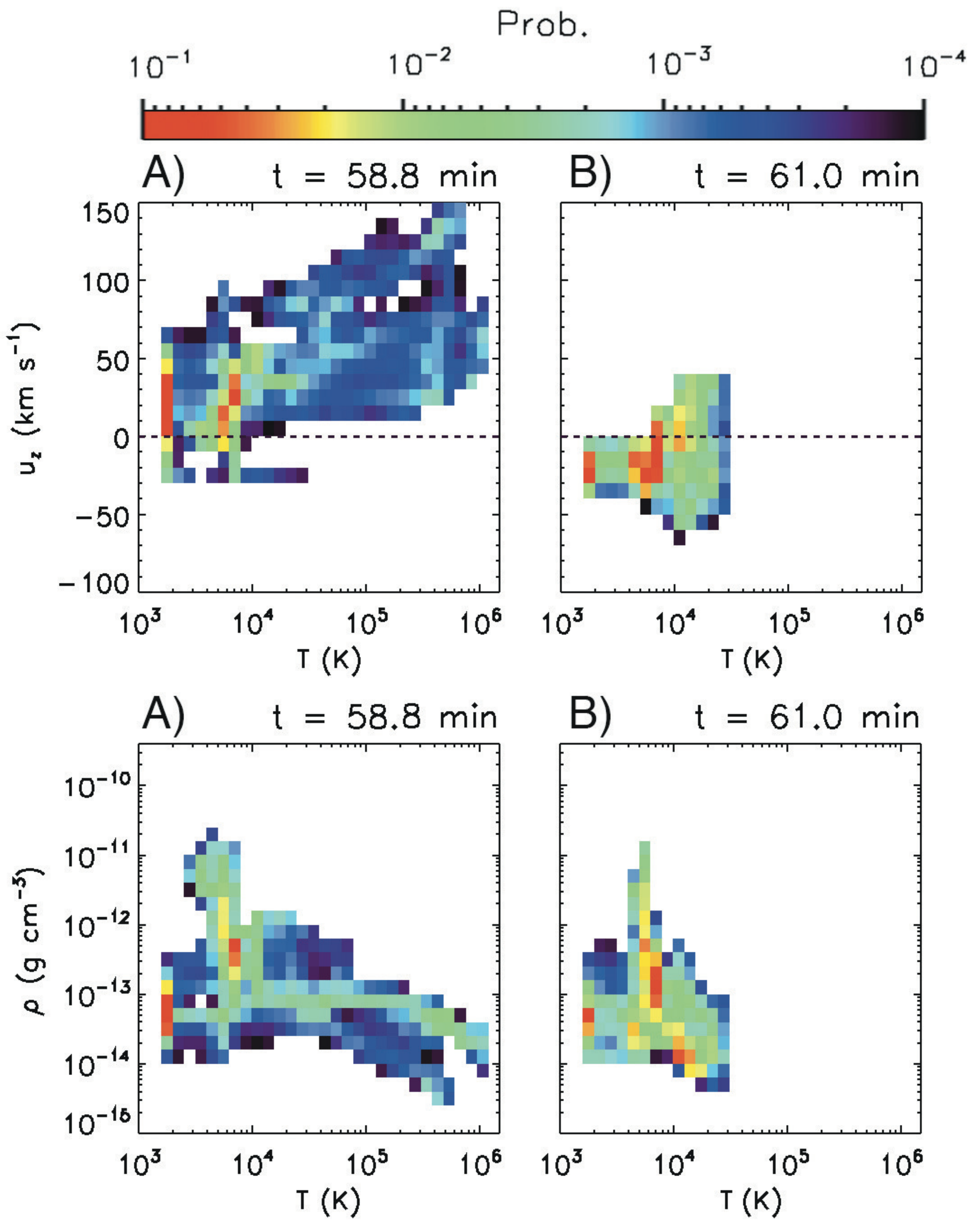}}
\caption{Double PDF plots for the temperature $\Tg$ and either the
  vertical velocity $u_z$ (upper panels), or the density $\rho$ (lower
  panels) of the Lagrangian elements. \label{velocidades}}
\end{figure}

\Needspace{5\baselineskip}
\subsection{Further properties: velocity, temperature and density}\label{sub:velocities}

We describe now some further properties of the ejecta:
velocities, temperatures and densities.
Figure~\ref{velocidades} contains double PDF plots for the
vertical velocity $u_z$ (upper row), and the density $\rho$
(lower row) versus the temperature, $\Tg$, of the Lagrangian
elements. The panels illustrate representative phases of the
surge: the launch phase (A panels); and the instant where
the surge reaches its maximum vertical extent (B panels).

\begin{itemize}

\item In the A panels we see that the majority of the plasma elements
  have cold temperatures (close to $2 \times 10^3$~K), densities
  around $10^{-13}$ g cm$^{-3}$, and velocities of a few tens of km
  s$^{-1}$: these are elements located in the dome at that
  time. Further,  there is a group of elements clustered at a
  temperature of $6$ to $7$ thousand K, possibly near the phase of
  hydrogen \ionizationrecombination, with small, positive velocities
  also of tens of km s$^{-1}$. This group contains a mixture of
  elements that have reached that temperature either through heating
  of cold plasma (Population \popstandard)  or
  cooling of hot plasma (Populations \popdometohot\ and \popcorona)
  through the action of the different entropy sources.
  Additionally, in the density panel there are two extended tails
  of elements toward higher temperatures and velocities. Those
  elements correspond to, on the one hand, hot, low density
  Population-\popcorona\ plasma originating in the corona, and, on the other hand, denser plasma from Population
  \popdometohot\ undergoing its heating-cooling phase. As we saw in
  Section \ref{sec:accel_launch}, the elements of Populations \popdometohot\ and
  \popcorona\ suffer the largest accelerations and, as a
  consequence, the range of velocities is between 20-150 km
  s$^{-1}$.

\item The B panels in the figure 
  are representative of the phase of maximum vertical
  development of the ejection. We see that basically the
  whole ensemble is already falling, albeit with small
  velocities ($|u_z| < 30$ km s$^{-1}$). Concerning the
  temperatures, there is an important concentration of
  particles at the temperatures of hydrogen
  \ionizationrecombination (around $6\times10^3$~K),
  \ionizationrecombination of \ion{He}{1}/\ion{He}{2} (around $\Tg \sim
  10^4$~K, label ``4'') and, to a lesser extent, of \ion{He}{2}/\ion{He}{3} 
  (around $\Tg \sim 2 \times 10^4$~K, label ``5'') -- see
  also the discussion about this issue in
  Section~\ref{sec:disc_limitations}. The density range for this phase
  of the cool surge is between $10^{-14}$ and
  $10^{-11}$~g~cm~$^{-3}$. In the later phases of the surge, the
  velocities continue in the range of a few to tens of km~s$^{-1}$. 

\end{itemize}

The resulting global picture of the surge during its
main development phase corresponds to plasma with velocities of tens
of km~s$^{-1}$.
The temperatures tend to be 
$0.6-1\times10^4$~K (but with a small population which have retained
their original cold temperature of a few thousand K) and the densities
are in a large range between $10^{-14}$ and $10^{-11}$ g cm$^{-3}$.

%

\Needspace{5\baselineskip}
\section{The detachment of the cool ejection from the dome}\label{sec:the_cleft} 
\begin{figure}
\epsscale{1.2}
\centerline{\plotone{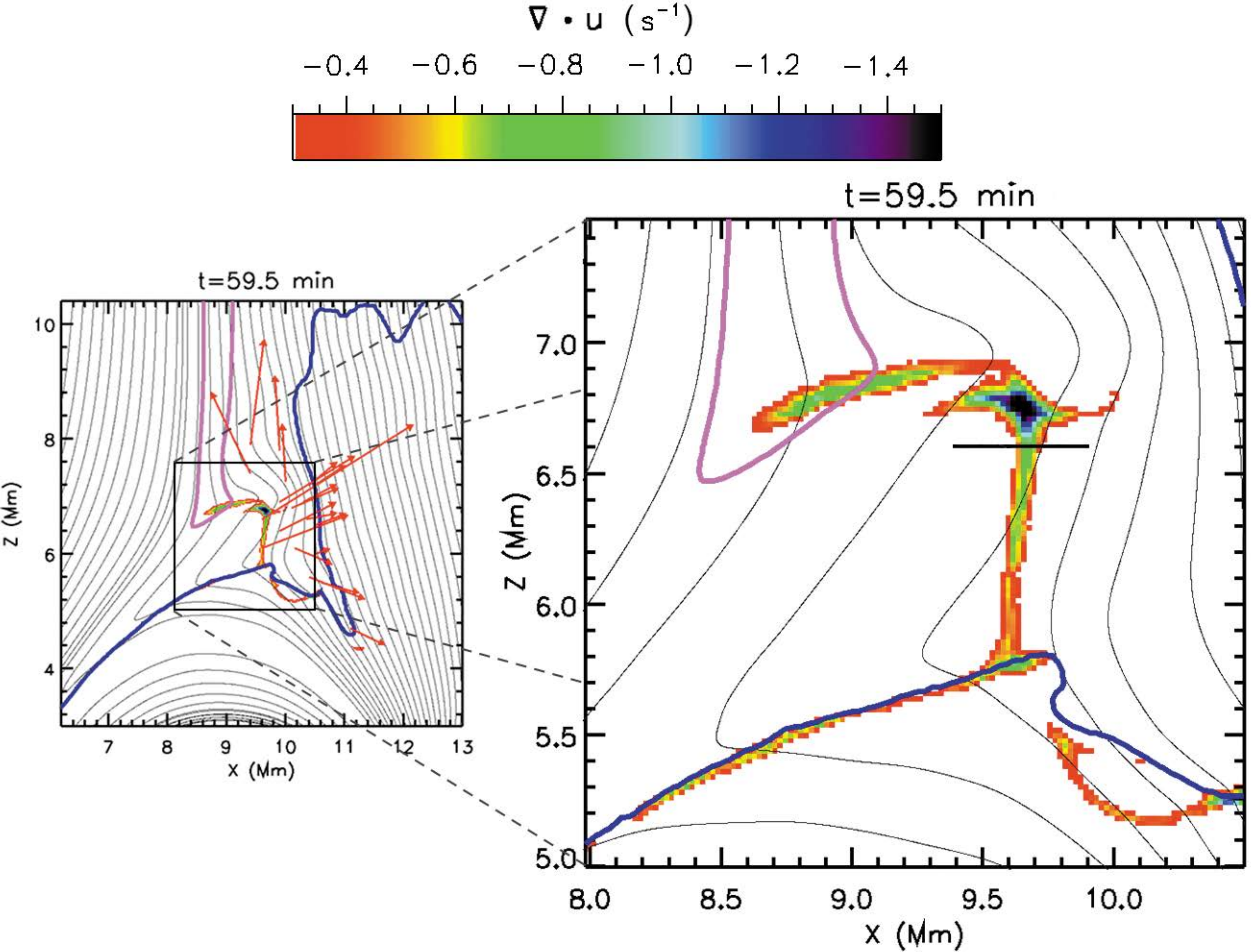}}
\caption{Left: Map of the velocity field divergence, $\nabla
  \cdot$\textbf{\textit{u}}. Only the pixels where $\nabla
  \cdot$\textbf{\textit{u}} $ < -0.5 $~s$^{-1}$ are shown in
  color. Pink and blue contours are the same that in
  Figure~\ref{wall}. The magnetic field is superimposed as
  black lines while the velocity field in the detachment
  region is shown with red arrows. Right: Zoom out for the
  previous panel to highlight the shock
  region. The horizontal black line is the cut used 
in Figure \ref{fastshock} to study the fast shock.\label{cleft}}
\end{figure}

Looking back at Figure~\ref{sec:cold_wall}, we realize that
from panel D onward the ejecta adopt the shape of a detached
{\it wall}, a cool and dense wall. Going a little earlier in
time (panel C), we locate the origin of the detachment in
the fact that the dome is being split in two at $x\approx
11$ Mm, the process taking place mainly between $z\approx 4$
and $z\approx 6$ Mm. The appearance of this {\it cleft} is
especially noticeable following the blue temperature
contour ($\Tg = 3 \times 10^4$~K). In
panels D and E, the detachment is seen to be
complete and the ejecta are from then on a separate
wall-like structure.

The explanation of this phenomenon lies in the formation of a series of
shocks above the dome starting at $t \approx 59$~min (check also the
density and temperature evolution shown in the accompanying Movie 1).
Successive blobs of plasma coming up from the reconnection site along the top
of the dome impinge on the surge. Strong shocks are created that deform and
redirect plasma in the blob, last for a brief period of time and then
weaken. A new 
blob arrives and creates again a shock system of the same kind.  To
illustrate the shock region in one of these collision events,
Figure~\ref{cleft} (left panel: general view; right-panel: blow-up of the
shock region) shows a map of the divergence of the velocity field, $\nabla
\cdot$\textbf{\textit{u}}, thus signposting the locations where a large
compression is taking place. Further, the figure contains the blue ($3 \times
10^4$~K) and pink ($1.2 \times 10^6$~K) isotemperature contours of
Figure~\ref{wall}, and a collection of field lines drawn as black
curves. Also, the arrows show the velocity field in the detachment region,
between the hot jet and the cool ejecta.

From the color map we see that the shock front has a wedge-like or arrowhead
shape, which is a common feature of the successive shocks seen during the
detachment phenomenon. The shocks cause high levels of compression and
heating of the plasma going through it. The two sections of the arrowhead
show distinctive features: the upper part, which is roughly horizontal and
nearly perpendicular to the field lines in the postshock region, resembles a
slow-mode shock almost of the switch-off kind. This could be related with the
slow-mode shocks generated when plasmoids collide with the ambient magnetic
field after being ejected, as illustrated by
\cite{YangL2013}. This shock is directly related with
  the hot jet: it is located at the base of the latter (see the pink
  contours) and the plasma goes through it before flowing along the horn-like
  jet field lines. We leave its study for a follow-up paper dealing with the
  properties of the hot jet.

\begin{figure}
\epsscale{1.20}
\centerline{\plotone{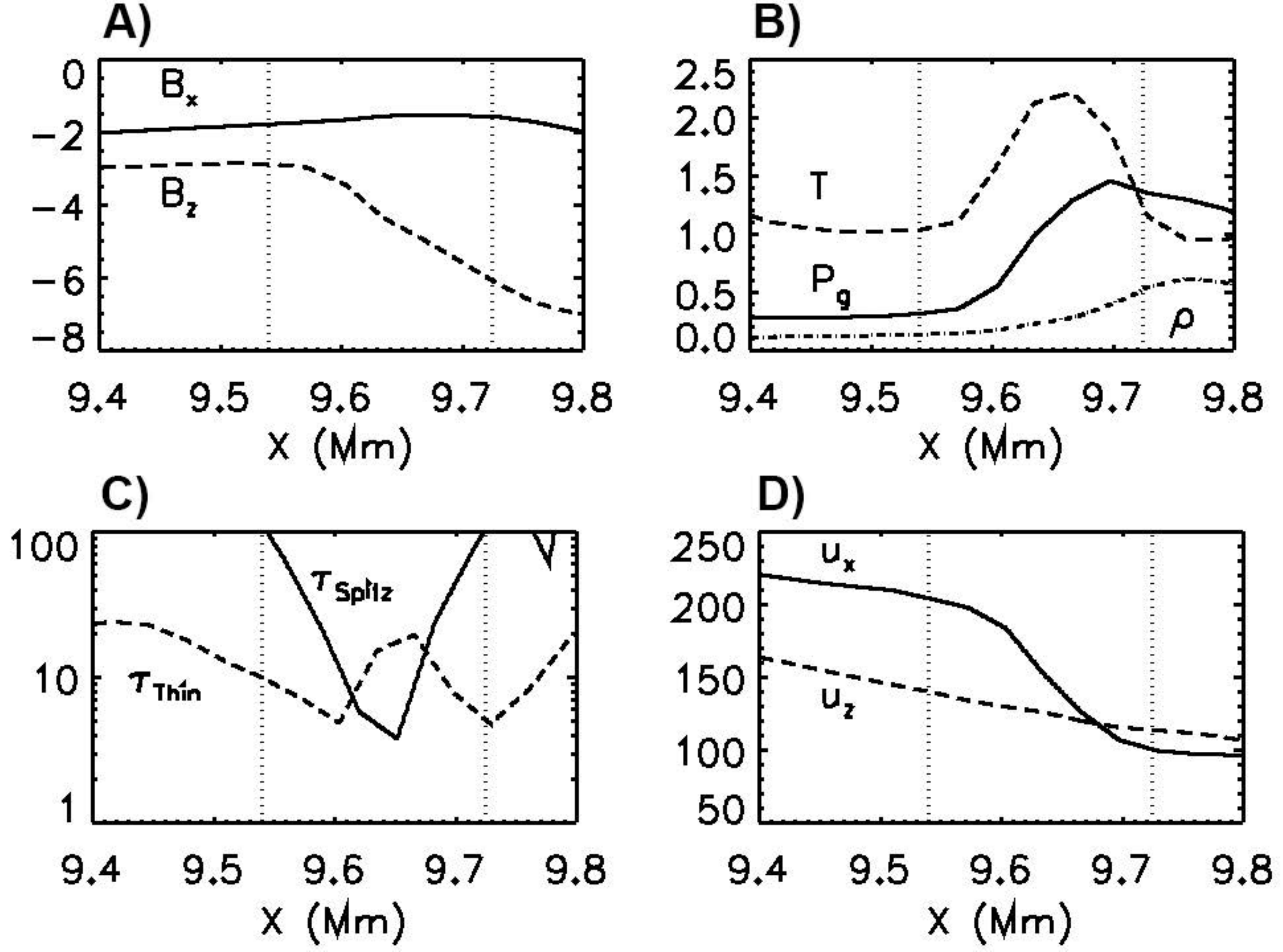}}
\caption{Jump relations along the horizontal black line
    in the right panel of Figure \ref{cleft}. The auxiliar vertical lines enclose
    the compression region where $\nabla \cdot$\textbf{\textit{u}} $\leq
    -0.2$ s$^{-1}$.  Panel A: Magnetic field $B_x$ and $B_z$ in G. Panel B:
    Temperature $T$ normalized to $4 \times 10^5$ K, gas pressure $P_g$ in
    erg cm$^{-3}$, and density $\rho$ normalized to $10^{-13}$ g
    cm$^{-3}$. Panel C: characteristic times for the Spitzer conductivity,
    $\tau_{Spitz}$, and the optically thin losses, $\tau_{Thin}$, in
    seconds. Panel D: velocities $u_x$ and $u_z$ in km
    s$^{-1}$. \label{fastshock}}
\end{figure}   

The lower, almost vertical branch of the wedge, in turn, is a shock directly
associated with the detachment process studied in this
section. The field lines cross it but subtending only a
  small angle to the tangent direction to the shock front. $\nabla
  \cdot$\textbf{\textit{u}} has high compression values of about
  $-0.8$~s$^{-1}$, sometimes reaching even $-3.0$~s$^{-1}$. Plasma traverses
  the structure from the left. Figure \ref{fastshock} shows the profiles
  across the shock for a number of relevant variables. To that end, we plot
  those variables along the horizontal black line plotted in the right panel
  of Figure \ref{cleft}.  The $B_z$ component (panel A) is not far from the
  perpendicular component of the field to the front normal. This component
  increases by a factor of 2 in absolute value across the shock, which
  suggests that the shock is a moderately strong, fast shock. This is also
  supported by the fact that the quasi-parallel component of the velocity
  ($u_z$, panel D) does not change substantially across the shock, whereas
  the normal velocity changes by a factor 2, approximately.  The temperature
  has a suggestive profile (panel B), that we can understand with the help of
  the entropy sources (panel C). In the first half of the shock the
  temperature increases, mostly because of the compression work experienced
  by the plasma element when entering the shock. In the center and final half
  of the shock, however, $T$ reaches a plateau and decreases: this is
  probably due to the action of the heat conduction (to a limited extent also
  of the optically thin radiation cooling): the characteristic cooling time
  scales are low ($4$ s for the former, see panel C), and fit with the
  duration of the transit of the plasma across the shock if one takes into
  account the motion of the front as a whole. Finally, the density increases
  by a factor $5$ (panel B), which is larger than the maximum allowed for
  adiabatic shocks: the large compression ratio is reached thanks to the
  entropy decrease due to the non-adiabatic effects.  As a further
  consequence of the heat conduction, the thermal energy is distributed
  efficiently along the individual field lines well beyond the shock itself,
  giving rise to the structure along the cleft that marks the boundaries of
  the cool ejection and lets it appear as a separate domain. The velocities
  involved in the shock, panel D, are on the order of a hundred km s$^{-1}$.
 

The plasma diverted downward after crossing the vertical section of the shock
penetrates deeper into the underlying dome as successive shock systems are
formed. When the last one in the series ends ($t \approx 61$~min, Panel F of
Figure~\ref{wall}), the surge is completely detached from the remnants of the
dome on the left. The hot plasma domains at that time have the classical
inverted-Y or Eiffel-tower shape commonly seen in observations, with one of
the legs of the tower coinciding with the {\it cleft}. Meanwhile, the cool
surge enters the decay phase following the swaying motion explained at the
beginning of Section \ref{sec:cold_wall}.  Both the hot and cool ejections
finally disappear almost simultaneously at around $t = 66$~min.

%

\section{Discussion}\label{sec:5}

We have performed a 2.5D radiative-MHD numerical experiment of emergence of
magnetized plasma through granular convection and into the atmosphere. The
time evolution of the system leads to the ejection of part of the emerged
material as a cool and dense surge. The experiment was done with the Bifrost
code, which includes a realistic multi-component equation of state as well as
modules for photospheric and chromospheric radiation transfer, heat
conduction and optically-thin radiative cooling in the corona. In the
following we first provide a comparison with observational data 
(Section \ref{sec:disc_observations}) and then discuss the relevance of some of
the entropy sources not included in the flux emergence experiments so far
(Section \ref{sec:disc_sources}). The final paragraphs point out a number of
limitations of the present experiment that may be overcome in the
future (Section \ref{sec:disc_limitations}).

%

\Needspace{5\baselineskip}
\subsection{Observations}\label{sec:disc_observations}
A first block of quantities that can be compared to observations concern the
size, timescale and kinematic properties of the surge. A more in-depth
comparison must be done through a-posteriori synthesis of different spectral
lines based on the numerical boxes. However, the most important spectral
lines that one could use for this comparison (like \Halpha, \CaII\ H+K or
\Heteneightthirty) require careful treatment including NLTE aspects; this
kind of approach must therefore be left for future work.

The height of the surge in our experiment varies considerably in the
different stages of the evolution. At the time of maximum development, the
ejecta constitute a vertically elongated object with height about $13$ Mm and
width about $2$ Mm. The observed length (see Section~\ref{sec:introduction})
falls typically in the interval $10$ - $50$ Mm, so the height of our surge is
within the observed range, even though toward its lower limit. This fits with
the fact that the experiment deals with a simple emergence event into a
coronal hole, whereas many classical observations refer to surges measured in
the context of flare episodes in active regions, which involve a larger
amount of magnetic flux and where larger structures should be expected. The
cool ejection in our experiment lasts for about $7 - 8$ min, which, again, is
toward the lower limit of the observed durations (several minutes to one
hour, \citealt{Jiang2007}, \citealt{Vargas2014}).  Regarding the velocities,
although high velocities of up to $150$ km s$^{-1}$ can be reached along the
launch phase, during most of the surge evolution the mass elements have
rising or falling velocities below $50$ km s$^{-1}$, and the ejection is not
collimated.  The observations, in turn, yield a velocity range of $10 - 200$
km s$^{-1}$, as inferred mainly from \Halpha\ measurements
(\citealp{roy1973}, \citealp{canfield1996}, \citealp{Chae1999},
\citealp{Jibben2004}, \citealp{Uddin2012}, \citealp{nelson2013},
\citealp{Vargas2014}, among others), which is compatible with the results of
the experiment.

Concerning the acceleration, in the experiment we have detected two different
patterns of behavior: (1) during the launch phase, the mass elements suffer
large accelerations, well in excess of solar gravity; (2) when near the apex
of their individual trajectories, the acceleration values are remarkably
close to $\grav$. There is no definitive observational value to use for a
comparison here: in the paper by \citet{roy1973}, the author reports a fast
rising phase for the surge with acceleration of $0.24 - 2.1$ km s$^{-2}$,
i.e., roughly $1 - 10\ \grav$. Observed values for the acceleration at the
time of maximum and in the decay phase are more difficult to obtain. In their
recent paper, \cite{nelson2013} detected an apparent parabolic trajectory for
the cool ejection in their study, but no particular value for the
acceleration was given.

%
\Needspace{5\baselineskip}
\subsection{The relevance of the entropy sources}\label{sec:disc_sources}

An adequate treatment of the entropy sources and sinks in the energy
equation, or, more generally, of the material properties of the plasma, like
its EOS, is important when studying the formation and time evolution of the
cool ejections. Thanks to the possibilities afforded by the
Bifost code and to an extensive Lagrange tracing of the mass elements of the
surge, we have been able to distinguish different patterns of behavior among
them and group them into separate populations. One of those populations,
Population \popdometohot, provides a good illustration in that sense. That
population covers $34$\% of the surge cross section at the time of maximum
development.  It reaches high temperatures, between $10^5$ and
$10^6$ K, typically when going through the current sheet, but is then brought
back down to classical surge temperatures of order $10^4$ K thanks to the
action of the radiation losses and thermal conduction terms. This population
could not be obtained in more idealized experiments, like those of
\citealp{Yokoyama:1996kx}, \citealp{Nishizuka:2008zl}, \citealp{jiang2012},
\citealp{Moreno-Insertis:2013aa}. The first authors, for instance, find that
the material in the surge structure is not heated significantly along its
life (which would roughly correspond to the behaviour of our populations
\popstandard\ and \popcold). Instead, we find that a non-small amount of the
plasma in the surge suffers heating/cooling processes that lead them to high
temperatures during a fraction of its life. This explains part of the
structural properties of the modeled surge and may also be of interest
concerning its detection.   

The importance of a proper treatment of the entropy sources and of the
equation of state may also apply to other cool ejections such as the
macrospicules. Chromospheric material and hotter, transition-region material
probably coexist in these objects, as indicated by their detection both in
\Halpha\ and in the EUV line \Hethreeofour. However, the numerical
experiments in the literature \citep[e.g.][]{Murawski2011,Kayshap2013} are of
the idealized kind, so, while possibly capturing various basic features of the
macrospicule phenomenon, they may also miss important aspects.

%

\Needspace{5\baselineskip}
\subsection{The progress toward realism in the theoretical modeling of surges
  following from flux emergence}\label{sec:disc_limitations} 

The essential component in the observed solar surge phenomenon is plasma with
chromospheric temperatures and densities, as follows from their detection in
spectral lines like \Halpha, \CaII\ H+K, \CaII\ 8542 $\angstrom$ or
\Heteneightthirty. Like for other important phenomena of the low solar
atmosphere (prominences are a prime example for this), their theoretical
study is intricate because of the difficulties of coping with the
material properties of the chromospheric plasma. All previous numerical
studies of surges following from flux emergence were done on the basis of
highly idealized models, without radiation transfer nor a multi-component
equation of state with realistic abundances, partial ionization processes,
etc.  Our present paper constitutes a large step forward in that direction,
given the degree of realism of the material modules of the Bifrost code, as
explained in Section~\ref{sec:bifrost}. In the following we first compare our
results with those of the 3D experiment of
\citet{Moreno-Insertis:2013aa}. Then, a few limitations of the present
experiment are discussed, namely the presence in flux emergence models of cool and
dense plasma domains in the low atmosphere, the effects of partial
ionization on Ohm's law and the lack of \ionizationrecombination
equilibrium in processes occurring on short timescales.

Our approach allowed us to gain new insights compared with previous idealized simulations of the ejection of cool surges, even with the recent 3D experiment by \citet{Moreno-Insertis:2013aa}.  Major differences between the two experiments come from the inclusion in our case of detailed material properties, radiation transfer and heat conduction, which have allowed us, e.g., to discern different plasma populations that later constitute the cool surge, or to study the initial interaction of the rising plasma with realistic granulation, or to identify the process of detachment and decay of the surge. In a 2D experiment one can reach much higher spatial resolution, which facilitates the study of many aspects difficult or impossible to consider in a 3D problem, like the formation and evolution of plasmoids or the shock structures associated with the jets. Finally, our Lagrange tracing has an extremely high cadence (thanks again to the reduced storage demands of a 2D experiment), and this is advantageous when pursuing the motion of the plasma elements across regions with strong gradients. On the other hand, various general properties of the surge in this paper are in agreement with the simulation of \citet{Moreno-Insertis:2013aa}: being three-dimensional, the cool ejecta in their experiment had the shape of an almost circular plasma wall with chromospheric density surrounding the emerged region, even if the largest concentration was found at the base of the hot jet. There, the cool domain had a height ($\sim 10$ Mm), similar to that obtained in the present 2.5D experiment, and width ($\sim 6 $ Mm), which is wider than in the present paper, perhaps because of the lack of realistic convection cells in their experiment, which can modify the horizontal sizes of the emerged structures. The surge velocities, around $50$ km s$^{-1}$, are also within the range given by those authors for their cool ejecta.

When large magnetized plasma domains rise from the solar interior to the low
atmosphere, a dense and cold plasma dome is formed, as repeatedly shown in
the numerical experiments since the 1990s. At the interface between the dome
and the overlying atmospheric material a large density gradient
arises. Numerical codes tend to smooth that sharp density contrast through
diffusion, in many cases via some explicit diffusion term, like in Bifrost,
or, in a less controllable fashion, through the hidden, intrinsic diffusion
of the numerical scheme, like in formally ideal MHD codes. Irrespective of
whether a process of mixing takes place in such interfaces in the actual Sun,
any result associated with this diffusion in the theoretical models must be
handled with care. In our case, we have identified a family of plasma
elements originating in coronal heights (population \popcorona,
Section~\ref{sec:corona}) whose density is increased to a large extent via
this kind of diffusion process when they pass near the current sheet before
being incorporated to the surge. The initial mass of this family is
negligible compared to the final mass of the surge; in some sense, that
family is swallowed by the much more dense material coming from the dome, so
the qualitative (and, to a large extent, quantitative) properties of the
final surge should be widely independent of the evolution of this particular
population.  A different issue concerns the dome itself: the large
expansion associated with the rise leads it to adopt cold temperatures, below
$2000$ K. The ad-hoc heating term mentioned in Section~\ref{sec:2} is then
activated in the calculation to prevent the plasma from cooling to lower
temperatures, for which the radiation tables used by Bifrost become inaccurate
\citep[see][]{Leenaarts:2011qy}. The material of the surge originates
essentially in the dome, so, in spite of the enormous advantages of the new
generation of MHD codes compared with the previous idealized models, a fully
realistic treatment of the evolution of the surge in its formation stage must
await the completion of material modules for the codes adequate 
to the very cold plasma volumes in the low atmosphere.

In the same vein, another aspect that must be improved in future models of
the solar surges is the use of a generalized Ohm's Law incorporating partial
ionization effects. On the basis of the general results of
\cite{Leake:2006kx}, \cite{Arber:2007yf}, \cite{Martinez-Sykora:2012uq,
  Martinez-Sykora2015} and \cite{Leake:2013dq}, among others, we expect that
these effects may allow some slippage of magnetic field and plasma via
ambipolar diffusion and counteract to some extent the cold temperatures of
the rising dome.  This could affect the populations obtained in the surge,
especially Population \popcold, see section~\ref{sec:popcold}. As a final
item in the list of limitations in the realism of the current model, we
mention here the lack of non-thermal equilibrium in the ionization/recombination
processes of hydrogen and helium. As already proposed long ago
\citep{kneer_1980}, in chromospheric processes that occur on comparatively
fast time scales (e.g., in shocks), the ionized species, especially hydrogen
and helium, may take longer to recombine than predicted by
local-thermodynamic-equilibrium (LTE) equations (see the recent results by
\citealt{Leenaarts:2007sf} and \citealt{golding2014}). This problem is
particularly important if one tries to obtain a posteriori, i.e., on the
basis of the calculated computational boxes, synthetic spectra for the
hydrogen or helium lines from plasma at temperatures around $6 \times 10^3$ K
(for H) or between $1$ and $2 \times 10^4$ K (for He). However, the time
evolution of the system itself may also be affected in a non-negligible way
by this departure of LTE. The inclusion of the non-equilibrium effects into
the models of surges is therefore another improvement that must be
incorporated in future extensions of the present work.

%

\ \vspace{-2mm} 
\acknowledgments We gratefully acknowledge financial support
by the Spanish Ministry of Economy and Competitiveness (MINECO) through
projects AYA2011-24808 and AYA2014-55078-P, as well as by NASA through grants
NNX11AN98G, NNM12AB40P and NNX14AI14G (HGCR grant) and contracts NNM07AA01C
(Hinode) and NNG09FA40C (IRIS). The authors thankfully acknowledge the
computer resources and the technical expertise and assistance provided at the
LaPalma supercomputer installation (IAC, Spain) and at the Teide
High-Performance Computing facilities (Instituto Tecnologico y de Energias
Renovables, ITER, Spain), where the calculations presented in this paper were
carried out.  Use for test runs of the Pleiades cluster through the computing
project s1061 from NASA's HEC division is also acknowledged. Finally, the
authors are grateful to the members of the Bifrost development team for their
help with the Bifrost code, and to the anonymous referee for his/her constructive comments.

%

\bibliographystyle{apj} \bibliography{collectionbib}

\end{document}